\documentclass[11pt,nobalance]{article}
\usepackage{lmodern}
\usepackage{amssymb,amsmath}
\usepackage{gensymb}

\usepackage{empheq}
\usepackage[usenames,dvipsnames]{color}
\usepackage{longtable,booktabs}
\usepackage{graphicx,grffile}
\usepackage[left=.75in, right=.75in, top=1in,bottom=1in]{geometry}
\usepackage{authblk}
\usepackage{tabularx}
\usepackage{stfloats}
\usepackage{pxfonts}
\usepackage{xspace}
\usepackage{setspace}
\usepackage{algorithm}
\usepackage{algpseudocode}
\usepackage{subcaption}
\usepackage{listings}

\usepackage{booktabs}
\usepackage{multirow}
\usepackage{longtable}
\usepackage{makecell}

\usepackage{enumitem}
\setlist{nosep}
\usepackage{makecell}
\usepackage{titlesec}
\usepackage{placeins} 
\usepackage{xr}
\usepackage{cite}

\providecommand{\keywords}[1]{\textbf{\textit{\noindent Keywords:\xspace}} #1}

\usepackage{fancyhdr}
\newcommand{\REM}[1]{}

\MakeRobust{\Call}
\algnewcommand{\LComment}[1]{\State \(\triangleright\) {\color{black} \textbf{#1}}}
\renewcommand{\Comment}[1]{\hfill $\triangleright$ {\color[rgb]{0.5,0.5,0.5} \textbf{#1}}}

\newcommand{\Wc}[0]{W_\mathrm{c}}
\newcommand{\Wd}[0]{W_\mathrm{d}}

\definecolor{mygreen}{rgb}{0,0.6,0}
\definecolor{mygray}{rgb}{0.5,0.5,0.5}
\definecolor{mymauve}{rgb}{0.58,0,0.82}

\lstdefinestyle{ASCII-table}{
  float,
  floatplacement=htbp!
}

\title{An Atomistic Fingerprint Algorithm for Learning \textit{Ab Initio} Molecular Force Fields}
\author[ ]{Yu-Hang Tang}
\author[ ]{Dongkun Zhang}
\author[ ]{George Em Karniadakis}
\affil[ ]{Division of Applied Mathematics, Brown University, Rhode Island, USA 02912}
\date{}

\begin{document}

\maketitle

\paragraph{Abstract}
Molecular fingerprints, i.e. feature vectors describing atomistic neighborhood configurations, is an important abstraction and a key ingredient for data-driven modeling of potential energy surface and interatomic force.
In this paper, we present the Density-Encoded Canonically Aligned Fingerprint (DECAF) fingerprint algorithm, which is robust and efficient, for fitting per-atom scalar and vector quantities.
The fingerprint is essentially a continuous density field formed through the superimposition of smoothing kernels centered on the atoms.
Rotational invariance of the fingerprint is achieved by aligning, for each fingerprint instance, the neighboring atoms onto a local canonical coordinate frame computed from a kernel minisum optimization procedure. We show that this approach is superior over PCA-based methods especially when the atomistic neighborhood is sparse and/or contains symmetry.
We propose that the `distance' between the density fields be measured using a volume integral of their pointwise difference. This can be efficiently computed using optimal quadrature rules, which only require discrete sampling at a small number of grid points. We also experiment on the choice of weight functions for constructing the density fields, and characterize their performance for fitting interatomic potentials.
The applicability of the fingerprint is demonstrated through a set of benchmark problems.

\vspace{1em}

\section{Introduction}\label{introduction}

Molecular Dynamics (MD) simulations have been widely used for studying atomistic systems, \textit{e.g.} proteins and catalysts, due to their ability to precisely capture transient events and to predict macroscopic properties from microscopic details~\cite{zhao2013mature,Lindorff-Larsen2016}.
In its most prevalent implementation, the trajectory of an atomistic system is integrated in time according to the Newton's law of motion using forces calculated as the negative gradient of a Hamiltonian, whose functional form and parameters are collectively referred to as a force field~\cite{Rappe1992a,Cornell1995,WilliamL.Jorgensen1996}.
Traditionally, the pairwise and many-body terms that comprise a force field are derived empirically by fitting to quantum mechanical calculations and experimental data.

Three properties directly relate to the applicability of a force field: accuracy, transferrability, and complexity~\cite{frenkel2001understanding,leach2001molecular}. Over the years, a large number of force fields have been developed, each carrying a particular emphasis over these three properties. However, the combinatorial complexity of atomistic systems can easily outpace force field development efforts, the difficulty of which explodes following the curse of dimensionality~\cite{Cheng2014}.
A deceptively simple system that can demonstrate the situation is water, a triatomic molecule with a well-characterized molecular structure.  In fact, all common water models, such as SPC-E, TIP3P, and TIP4P, have only succeeded in reproducing a small number of structural and dynamical properties of water due to the difficulty in modeling strong intermolecular many-body effects such as hydrogen bonding and polarization~\cite{Braun2014,Boonstra2016}.

In lieu of a force field, quantum mechanical (QM) calculations can be employed straightforwardly to drive molecular dynamics simulations. The method achieves significantly better accuracy and transferrability by solving for the electronic structure of the system. However, the computational complexity of QM methods is at least cubic in the number of electrons, and consequently the time and length scales accessible by QM-driven molecular dynamics are severely constrained.

\begin{figure}[t!]
	\centering
	\includegraphics[width=0.9\columnwidth]{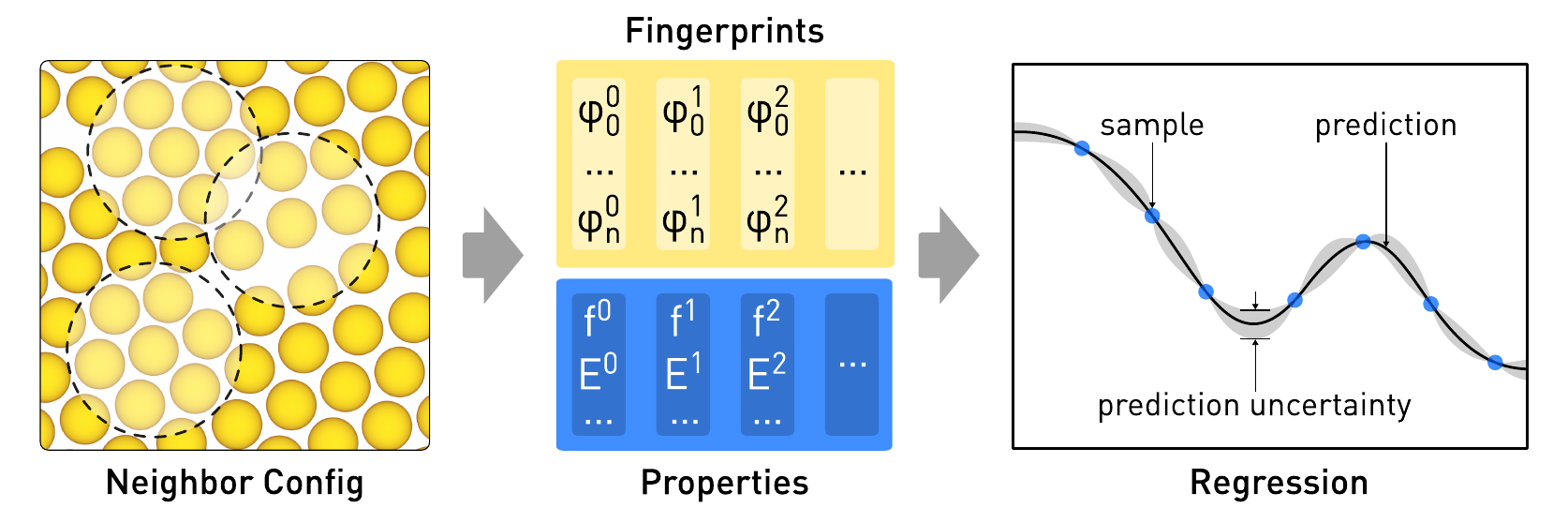}
	\caption{In the pipeline of machine learning-driven molecular computations, atomistic neighborhood configurations are transformed into feature vectors, called fingerprints, and used to train non-linear regression models.\label{fig:overview}}
\end{figure}

Assuming that there is smoothness in the potential energy surface of the atomistic system, one possible strategy to accelerate QM-driven molecular dynamics is to use QM calculations on only a subset of the time steps, and to interpolate for similar atomic configurations~\cite{Behler2011,Bartok2013a,Li2015c,Khorshidi2016}. A schematic overview of the process is given in Figure~\ref{fig:overview}, which is enabled by the recent development of high-dimensional nonlinear statistical learning and regression techniques such as Gaussian process regression~\cite{Rasmussen2006a} and artificial neural networks~\cite{Specht1991}.

This paper focuses on a particular aspect of the machine-learning-driven molecular computation pipeline, \textit{i.e.} fingerprint algorithms, whose importance arises naturally from the aforementioned regression protocol. A fingerprint is an encoding of an atomistic configuration that can facilitate regression tasks such as similarity comparison across structures consisting of variable numbers of atoms and elements. As has been pointed out previously~\cite{Bartok2013a}, a good fingerprint should possess the following properties:
\begin{enumerate}
\item It can be encoded as a fixed-length vector so as to facilitate regression (particularly for artificial neural networks).
\item It is \emph{complete}, i.e. different atomistic neighborhood configurations lead to different fingerprints and vice versa, and the `distance' between the fingerprints should be proportional to the intrinsic difference between the atomistic neighborhood configurations.
\item It is \emph{continuous} with regard to atomistic coordinates, and the change in fingerprint should be approximately proportional to the structural variation as characterized by, for example, some internal coordinates.
\item It is \emph{invariant} under permutation, rotation, and translation.
\item It is computationally feasible and straightforward to implement.
\end{enumerate}

Before we proceed to the details of our work, we will first briefly review several fingerprints that are closely related to our work, \textit{i.e.} the Smooth Overlap of Atomic Positions (SOAP) kernel~\cite{Bartok2013a}, the Coulomb matrix~\cite{Rupp2012}, and the Graph Approximated Energy (GRAPE) kernel~\cite{Ferre2017}.

\paragraph{Smooth Overlap of Atomic Positions (SOAP):}

The SOAP kernel is built on the idea of representing atomistic neighborhoods as smoothed density fields using Gaussian kernels each centered at a neighbor atom. Similarity is measured as the inner product between density fields, while rotational invariance is achieved by integrating over all possible 3D rotations, which can be performed analytically using the power spectrum of the density field. In fact, our fingerprint algorithm is inspired by this idea of treating atoms as smoothed density fields. However, we take a different approach to endorse the fingerprint with rotational invariance, and use the Euclidean distance instead of inner product as a distance metric.

\paragraph{Coulomb Matrix:}

The practice of using graphs to represent atomistic neighbor configurations was first implied by the Coulomb matrix, and later further formulated in the GRAPE kernel, where the \emph{diffusion distance} was proposed as a similarity measure between different local chemical environments \cite{Sun2014}. The idea is to construct an undirected, unlabeled graph $G = (\,V\,,\,E\,)$ with atoms serving as the vertices and pairwise interactions weighting the edges. For example, the Coulomb matrix can be treated as a physically-inspired Laplacian matrix \cite{Coifman2005}
\begin{align}
\mathbf{M} &= \mathbf{D} - \mathbf{A}\\
\mathbf{D}_{IJ} &=
  \begin{cases}
    0.5\,Z_I^{2.4} & \text{if } I=J\\
    0 & \text{if } I \neq J
  \end{cases}\\
\mathbf{A}_{IJ} &=
  \begin{cases}
    0 & \text{if } I=J\\
    -\frac{Z_I\,Z_J}{\lVert \mathbf{R}_I - \mathbf{R}_I \rVert} & \text{if } I \neq J
  \end{cases}
\end{align}
where the degree matrix $\mathbf{D}$ encodes a polynomial fit of atomic energies to the nuclear charge, while the adjacency matrix $\mathbf{A}$ corresponds to the Coulombic interactions between all pairs of atoms. Due to the use of only relative positions between atoms in the adjacency matrix, the Coulomb matrix is automatically invariant under translation and rotation. However, the matrix itself is not invariant under permutation, as swapping the order of two atoms will result in an exchange of the corresponding columns and the rows. To address this, the sorted list of eigenvalues of the Coulomb matrix can be used instead as a feature vector, while an $\ell_p$ norm can be used as a distance metric. In practice, due to the fact that the number of neighbor atoms may change, the shorter eigenvalue list is padded with zeros in a distance computation.

\paragraph{Graph Approximated Energy (GRAPE):}

The GRAPE kernel evaluates the simultaneous random walks on the direct product of the two graphs representing two atomistic neighborhood configurations. Permutational invariance is achieved by choosing a uniform starting and stopping distribution across nodes of both graphs. However, the cost of distance computation between two graphs scales as $\mathcal{O}(N^2)$ with a one-time per-graph diagonalization cost of $\mathcal{O}(N^3)$.

In the sections below, we present our new fingerprint algorithm, namely the Density-Encoded Canonically Aligned Fingerprint (DECAF). The paper is organized as follows: in Section~\ref{sec:cframe}, we introduce a robust algorithm that can determine canonical coordinate frames for obtaining symmetry-invariant projections; in Section~\ref{sec:density-field}, we present numerical recipes to use smoothed atomistic density fields as a fingerprint for molecular configuration; in Section~\ref{sec:demonstration}, we demonstrate the capability of the fingerprint via examples involving the regression of atomistic potential energy surfaces; in Section~\ref{sec:discussion}, we discuss the connection between our algorithm and previously proposed ones; we conclude with a discussion in Section~\ref{sec:conclusion}.

\section{Localized Canonical Coordinate Frame for Rotationally Invariant Description of Atomistic Neighborhood}
\label{sec:cframe}

\subsection{Kernel Minisum Approach}

To improve model generalization while minimizing data redundancy, a fingerprint algorithm should be able to recognize atomistic structures that differ only by a rigid-body transformation or a permutation of atoms of the same element, and to extract feature vectors invariant under these transformations. As summarized in Table~\ref{table:invariance}, a variety of strategies have been successfully employed by common fingerprint algorithms to achieve rotational invariance.

However, these approaches do not provide a means for the acquisition of vector-valued quantities in a rotational invariant form. One approach is to only acquire and interpolate the potential energy, a scalar quantity, and then take the derivative of the regression model. This approach, however, triggers the need for methods to decompose the total energy among the atoms, which is a property of the entire system rather than individual atoms~\cite{Bartok2010}.

Another approach proposed by Li \textit{et al.} \cite{Li2015c,Botu2015} is to project vector quantities onto a potentially overcomplete set of non-orthogonal basis vectors obtained from a weighted sum of the atomic coordinate vectors:
\begin{align}
\mathbf{V}_k = \sum_i \mathbf{x}_i\,\exp\left[-\left( \frac{ \lVert \mathbf{x}_i \rVert }{ R_\mathrm{c} } \right)^{p_k}\right].
\end{align}
However, the approach may suffer from robustness issues. For example, all of the $\mathbf{V}_k$ generated with different $p_k$ will point in the same direction if the radial distance of the atoms are all equal. Further, the configuration with 4 atoms at $(r \cos\varepsilon, r \sin\varepsilon),(0,r),(-r,0),(0,-r)$ leads to
\begin{align}
\mathbf{V}_k
&= c\cdot[ (r \cos\varepsilon, r \sin\varepsilon) + (0,r) + (-r,0) + (0,-r) ]\\
&= c\cdot r\cdot( 1 - \cos\varepsilon, \sin\varepsilon).
\end{align}
Thus, if $\varepsilon$ gets close to zero, $V_k$ will always point toward either $(0,1)$ or $(0,-1)$, even if the vector quantity of interest may point in other directions.

\begin{table}[htbp!]
\centering
\caption{Comparison of strategies used by fingerprint algorithms to obtain feature vectors which are invariant under translation, permutation, and rotation.}
\label{table:invariance}
\begin{tabular}{@{}llll@{}}
\toprule
\multirow{2}{*}{\textbf{Fingerprint}} & \multicolumn{3}{c}{\textbf{Invariance}}                                      \\ \cmidrule(l){2-4}
                                      & \textbf{Translation} & \textbf{Permutation} & \textbf{Rotation}              \\ 
\midrule
Coulomb matrix\cite{Rupp2012}         & relative distance    & sorting eigenvalues  & all vs. all graph              \\
Behler\cite{Behler2011}               & relative distance    & summation            & ignoring angular information   \\
SOAP\cite{Bartok2013a}                & relative distance    & summation            & integrating over all rotations \\
GRAPE\cite{Ferre2017}                 & relative distance    & uniform distribution & uniform distribution           \\ \bottomrule
\end{tabular}
\end{table}

Here, we present a robust kernel PCA-inspired algorithm for the explicit determination of a canonical coordinate frame, within which the projection of the atomistic neighborhood is invariant under rigid-body rotation. Furthermore, the canonical coordinate frame can be directly used to capture vector-valued quantities in a rotational-invariant form. Given $N$ atoms with position $\mathbf{x}_1,\ \ldots,\ \mathbf{x}_N \in \mathbb{R}^d$, we first formulate the $L_p$ PCA algorithm as an optimization problem where we seek a unit vector $\mathbf{w}^*$ that maximizes the sum of the projections:
\begin{align}
\mathbf{w}^* &= \mathrm{argmax}_{\lVert \mathbf{w} \rVert = 1 } \sum_{i=1}^{N} \left| \mathbf{w}^\mathsf{T} \mathbf{x}_i \right|^p\\
             &= \mathrm{argmax}_{\lVert \mathbf{w} \rVert = 1 } \sum_{i=1}^{N} \left| r_i \right|^p \left| \mathbf{w}^\mathsf{T} \mathbf{e}_i \right|^p,
\end{align}
where $r_i = \lVert \mathbf{x}_i \rVert$ is the distance from the origin to atom $i$, $\mathbf{e}_i = \mathbf{x}_i / r_i$ is the unit vector pointing toward atom $i$, respectively.
The optimization process can only uniquely determine the orientation of a projection vector up to a line, because $\left| \mathbf{w}^\mathsf{T} \mathbf{e}
\right| \equiv \left| -\mathbf{w}^\mathsf{T} \mathbf{e}
\right|$. As a consequence, further heuristics are needed to identify a specific direction for the PCA vectors.

To overcome this difficulty, we generalize the $\left| r_i \right|^p$ term into a weight function $g(r)$ of radial distance and the $\left| \mathbf{w}^\mathsf{T} \mathbf{e}_i \right|^p$ term into a bivariate kernel function $\kappa(\mathbf{w},\mathbf{e})$ between two vectors. We then attempt to seek a unit vector $\mathbf{w}^*$ that minimizes the kernel summation:
\begin{align}
\mathbf{w}^* = \mathrm{argmin}_{\lVert \mathbf{w} \rVert = 1 } \sum_{i=1}^{N} g(r_i)\,\kappa( \mathbf{w}, \mathbf{e}_i ).
\end{align}

\begin{figure}[htb!]
	\centering
	\includegraphics[width=0.95\columnwidth]{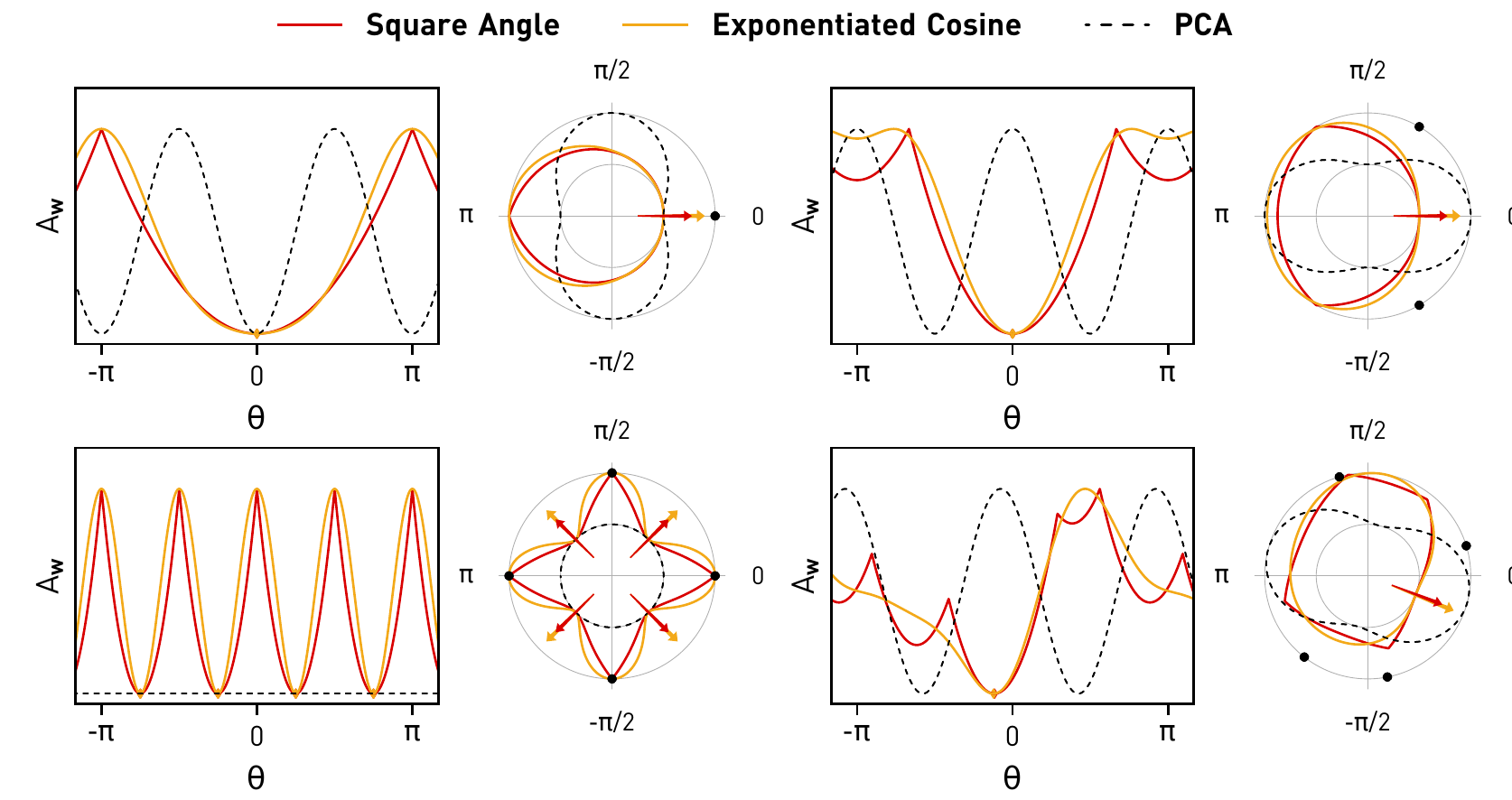}
	\caption{Shown here is an illustration of the minisum algorithm that determines projection vectors for rotational invariant description of atomistic neighbor configurations. Black dots represent atoms which all carry equal importance. The vectors that point from the origin to the atoms are used as the input to bivariate kernels to compute the minisum objective function, which are drawn in solid lines. For reference, the negated values of the PCA objective function are drawn in dashed lines. Projection vectors are obtained by finding the unit vector $\mathbf{w}^*$ that minimizes the objective function. \label{fig:minisum}}
\end{figure}

In particular, we have found a square angle (SA) kernel and an exponentiated cosine (EC) kernel that perform well in practice:
\begin{align}
\kappa_\mathrm{SA}(\mathbf{w},\mathbf{e}) &\doteq  \frac{1}{2} \arccos^2( \mathbf{w}^\mathsf{T} \mathbf{e} ), \\
\kappa_\mathrm{EC}(\mathbf{w},\mathbf{e}) &\doteq  \exp{\left(-\mathbf{w}^\mathsf{T} \mathbf{e}\right)}.
\end{align}
As shown in Figure~\ref{fig:minisum}, both kernels are minimal when $\mathbf{w}$ and $\mathbf{e}$ are parallel, and monotonically reach maximum when $\mathbf{w}$ and $\mathbf{e}$ are antiparallel. Intuitively, optimizing the minisum objective function generated by the SA kernel will yield a vector that, loosely speaking, bisects the sector occupied by the atoms. The EC kernel exhibits very similar behavior but leads to a smoother objective function.
As shown in Figure~\ref{fig:minisum}, this allows for the determination of a projection vector without ambiguity, even if the atom configuration contains perfect symmetry.

\begin{figure}[htb!]
	\centering
	\includegraphics[width=0.9\columnwidth]{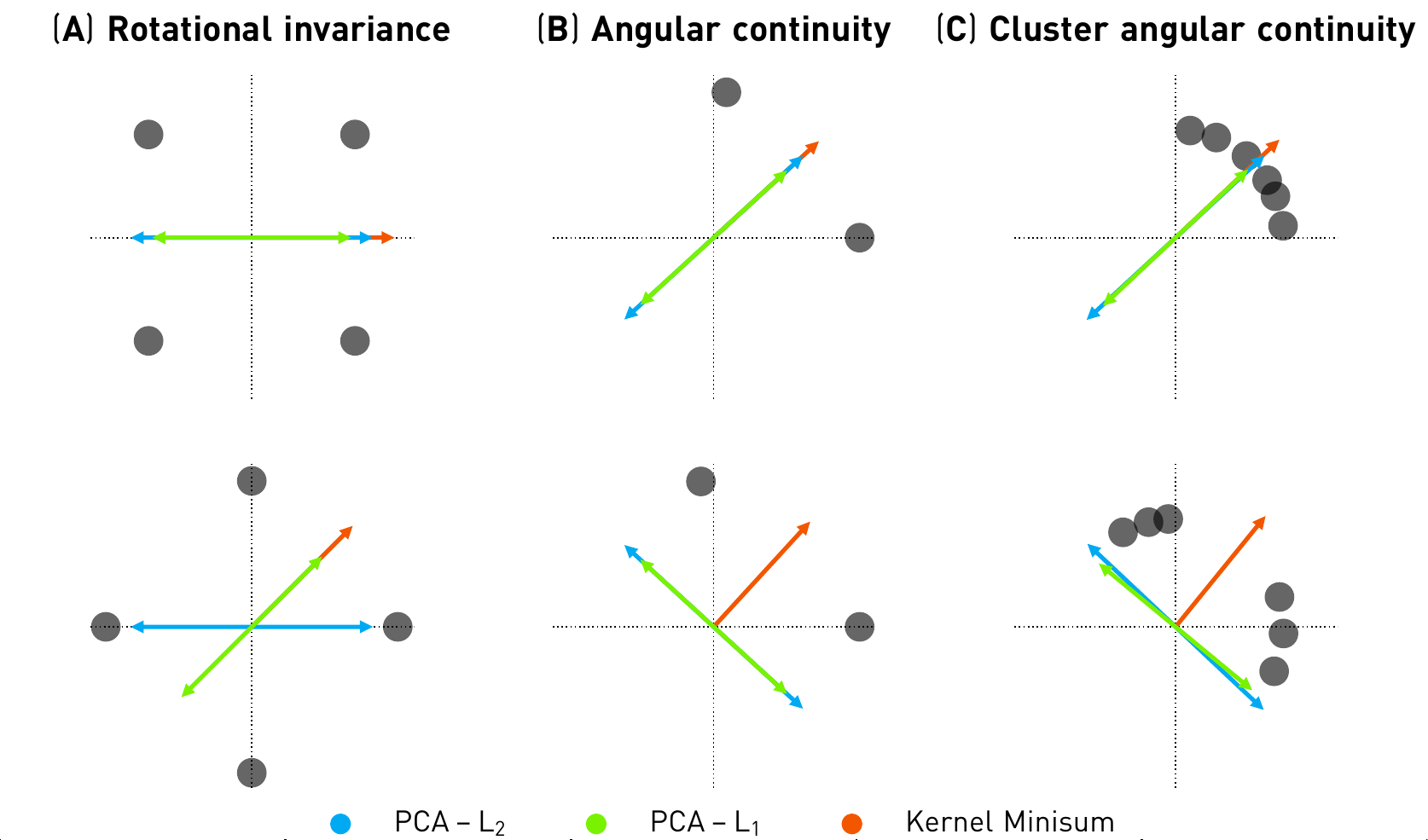}
	\caption{A comparison of the orthogonal bases obtained using principal components analysis (PCA) and kernel minisum with the square-angle (MSA) kernel. \textbf{(A)} The PCA algorithm, used in conjunction with the $L_2$ norm, fails to extract a principal axis that rotates with a system that exhibits planar $\mathrm{C}_4$ symmetry. Both MSA and PCA with the $L_1$ norm can accommodate this scenario. \textbf(B) Both $L_1$ and $L_2$ principal axes change orientation abruptly when the system undergoes a slight angular motion. In contrast, the MSA output is continuous with regard to this movement as it always bisects the angle formed by the atoms and the origin. \textbf{(C)} Loosely speaking, the MSA axis points at the majority direction of the atoms if only a single cluster is present within the cutoff distance, but bisects the angle between two atom clusters. This is different from PCA-based results and should deliver robust rotational invariance as well as continuity. Arrows are drawn at different lengths to improve visual clarity in situations of overlapping. They are to be understood as unit vectors. De-trending is not performed because the radial distances of the atoms carry physically significant information. \label{fig:msa}}
\end{figure}

A major advantage of the kernel minisum approach versus $L^p$ norm-based PCA, lies in its 1) robustness in the presence of structural symmetry; and 2) continuity of the resulting principal axes with respect to angular movement of the input data. As shown in Figure~\ref{fig:msa}, kernel minisum is particularly suitable for atomistic systems where strong symmetries are common and the continuity against angular movement is desired. The minisum framework can also be used with other customized kernels to suit for the characteristics of specific application scenarios.

\subsection{Solving the Kernel Minisum Optimization Problems}

The optimization problem can be solved very efficiently using a gradient descent algorithm as detailed below.

\paragraph{Square Angle:}
The objective function of the minisum problem using the square angle (SA) kernel is
\begin{align}
A_\mathrm{SA}(\mathbf{w}) \doteq \frac{1}{2} \sum_{i=1}^{N} g(r_i)\,\arccos^2( \mathbf{w}^\mathsf{T} \mathbf{e}_i ).
\end{align}
The gradient of $A_\mathrm{SA}$ with respect to $\mathbf{w}$ is
\begin{align}
\nabla_\mathbf{w}\,A_\mathrm{SA} = \sum_{i=1}^{N} -g(r_i)\,\frac{ \arccos( \mathbf{w}^\mathsf{T} \mathbf{e}_i ) }{ \sqrt{ 1 - (\mathbf{w}^\mathsf{T} \mathbf{e}_i)^2 } }\,\mathbf{e}_i. \label{eq:SA-derivative}
\end{align}
Note that $ \frac{ \arccos( \mathbf{w}^\mathsf{T} \mathbf{e}_i ) }{ \sqrt{ 1 - (\mathbf{w}^\mathsf{T} \mathbf{e}_i)^2 } } $ is singular when $\mathbf{w} \parallel \mathbf{e}_i$. This can be treated numerically by replacing the removable singularities at $ \mathbf{w}^\mathsf{T} \mathbf{e_i} = 1$ with the left-limit $ \lim_{\mathbf{w}^\mathsf{T} \mathbf{e}_i \rightarrow 1^-} \frac{ \arccos( \mathbf{w}^\mathsf{T} \mathbf{e}_i ) }{ \sqrt{ 1 - (\mathbf{w}^\mathsf{T} \mathbf{e}_i)^2 } } = 1$, while truncating the gradient at a finite threshold near the poles at $ \mathbf{w}^\mathsf{T} \mathbf{e_i} = -1$.

A local minimum can be iteratively searched for with gradient descent
while renormalizing $\mathbf{w}$ after each iteration. Moreover, due to the locally quadratic nature of the objective function, we have found that the Barzilai-Borwein algorithm \cite{BARZILAI1988} can significantly accelerate the convergence at a minimal cost. The algorithm is presented in Alg.~\ref{alg:MSA}.

\begin{algorithm}[h!]
\caption{Gradient descent for solving the square angle minisum problem.}
\label{alg:MSA}
\begin{algorithmic}[1]
\Function{MinSquareAngle}{$\mathbf{E} = [\mathbf{e}_1, \mathbf{e}_2, \ldots, \mathbf{e}_N]$, $\mathbf{G} = [g_1, g_2, \ldots, g_N]$, $\mathbf{w}_0$ }
\State $\mathbf{w} \gets \mathbf{w}_0$
\Repeat
\State Compute gradient $\nabla_\mathbf{w}$ using Eq.~\ref{eq:SA-derivative}.
\State Obtain tangential component of gradient $\nabla_\mathbf{w}^{\perp} \gets (\mathbf{I} - \mathbf{w}\mathbf{w}^\mathsf{T})\nabla_\mathbf{w}$
\If {at step 0}
  \State $\alpha \gets 0.01$ \Comment{Small initial step size for bootstrapping}
\Else
  \State $\alpha \gets \frac{ (\mathbf{w} - \mathbf{w}_{-1})^\mathsf{T} (\nabla_\mathbf{w}^{\perp} - \nabla_{\mathbf{w}_{-1}}^{\perp} ) }{ \lVert \nabla_\mathbf{w}^{\perp} - \nabla_{\mathbf{w}_{-1}}^{\perp} \rVert^2 }$ \Comment{Adaptive subsequent steps by Barzilai-Borwein}
\EndIf
\State Save $\mathbf{w}$ as $\mathbf{w}_{-1}$, save $\nabla_\mathbf{w}^{\perp}$ as $\nabla_{\mathbf{w}_{-1}}^{\perp}$
\State Update $\mathbf{w} \gets \mathbf{w} - \alpha\,\nabla_\mathbf{w}^{\perp}$ and normalize to unit length
\Until $\lVert \mathbf{w} - \mathbf{w}_{-1} \rVert < \varepsilon$
\State \Return $\mathbf{w}$
\EndFunction
\end{algorithmic}
\end{algorithm}

\paragraph{Exponentiated Cosine:}

The objective function of the minisum problem using the exponentiated cosine (EC) kernel is:
\begin{align}
A_\mathrm{EC}(\mathbf{w}) \doteq \sum_{i=1}^{N} g(r_i)\,\exp{\left(-\mathbf{w}^\mathsf{T} \mathbf{e}_i\right)}.
\end{align}
The gradient of $A_\mathrm{EC}$ with respect to $\mathbf{w}$ is
\begin{align}
\nabla_\mathbf{w}\,A_\mathrm{EC} = \sum_{i=1}^{N} -g(r_i)\,\exp{\left(-\mathbf{w}^\mathsf{T} \mathbf{e}_i\right)}\,\mathbf{e}_i. \label{eq:EC-derivative}
\end{align}
The gradient contains no singularity. However, it is not always locally quadratic or convex. This can cause the Barzilai-Borwein algorithm to generate negative step sizes and consequently divert the search towards a maximum. Luckily, this can be easily overcome by always using the absolute value of the step size generated by the Barzilai-Borwein algorithm. Such enforcement prevents the minimization algorithm from going uphill. The complete algorithm is given in Alg.~\ref{alg:MEC}.

\begin{algorithm}[h!]
\caption{Gradient descent for solving the exponentiated cosine minisum problem.}
\label{alg:MEC}
\begin{algorithmic}[1]
\Function{MinExpCosine}{$\mathbf{E} = [\mathbf{e}_1, \mathbf{e}_2, \ldots, \mathbf{e}_N]$, $\mathbf{G} = [g_1, g_2, \ldots, g_N]$, $\mathbf{w}_0$ }
\State $\mathbf{w} \gets \mathbf{w}_0$
\Repeat
\State Compute gradient $\nabla_\mathbf{w}$ using Eq.~\ref{eq:EC-derivative}.
\State Obtain tangential component of gradient $\nabla_\mathbf{w}^{\perp} \gets (\mathbf{I} - \mathbf{w}\mathbf{w}^\mathsf{T})\nabla_\mathbf{w}$
\If {at step 0}
  \State $\alpha \gets 0.01$ \Comment{Small initial step size for bootstrapping}
\Else
  \State $\alpha \gets \left| \frac{ (\mathbf{w} - \mathbf{w}_{-1})^\mathsf{T} (\nabla_\mathbf{w}^{\perp} - \nabla_{\mathbf{w}_{-1}}^{\perp} ) }{ \lVert \nabla_\mathbf{w}^{\perp} - \nabla_{\mathbf{w}_{-1}}^{\perp} \rVert^2 } \right| $ \Comment{Adaptive subsequent steps by Barzilai-Borwein}
\EndIf
\State Save $\mathbf{w}$ as $\mathbf{w}_{-1}$, save $\nabla_\mathbf{w}^{\perp}$ as $\nabla_{\mathbf{w}_{-1}}^{\perp}$
\State Update $\mathbf{w} \gets \mathbf{w} - \alpha\,\nabla_\mathbf{w}^{\perp}$ and normalize to unit length
\Until $\lVert \mathbf{w} - \mathbf{w}_{-1} \rVert < \varepsilon$
\State \Return $\mathbf{w}$
\EndFunction
\end{algorithmic}
\end{algorithm}

As shown in Table~\ref{table:minisum-convergence}, both Alg.~\ref{alg:MSA} and Alg.~\ref{alg:MEC} converge quickly and consistently across a wide range of representative point configurations commonly found in molecular systems. However, the gradient descent method can only find local optima. Thus, multiple trials should be performed using different initial guesses to ensure that a global minimum can be located.

\begin{table}[htbp!]
\centering
\caption{Listed here is the number of iterations and initial guesses used by the gradient descent algorithm to find an local optimum of the kernel minisum problems. The numbers are averaged over 500 repetitions, and the convergence criterion is $10^{-14}$. In cases where the iterative algorithm does not converge within 64 iterations, the optimization will be restarted with a new guess.}
\label{table:minisum-convergence}
\begin{tabular}{@{}llllll@{}}
\toprule
\textbf{Kernel}              & \multicolumn{2}{c}{\textbf{Square Angle}} &  & \multicolumn{2}{c}{\textbf{Exponentiated Cosine}} \\ \cmidrule(r){1-1} \cmidrule(lr){2-3} \cmidrule(l){5-6}
\textbf{Configuration}       & Itrs./Guess            & Guesses          &  & Itrs./Guess                & Guesses              \\ \midrule
Single point                 & 8.8                    & 1                &  & 7.9                        & 1                    \\
Two points, angle $< \pi/2$   & 7.1                    & 1                &  & 7.7                        & 1                    \\
Two points, angle $\ge \pi/2$ & 6.1                    & 1                &  & 6.3                        & 1                    \\
Two points, angle $= \pi$     & 6.0                    & 1                &  & 5.6                        & 1                    \\
Planar $C_3$                 & 6.2                    & 1                &  & 6.3                        & 1                    \\
Planar $C_4$                 & 5.6                    & 1                &  & 6.9                        & 1                    \\
Tetrahedra                   & 10.6                   & 1                &  & 7.7                        & 1                    \\
Octahedra                    & 11.7                   & 1                &  & 9.4                        & 1                    \\
Improper $S_4$               & 17.9                   & 1                &  & 23.1                       & 1                    \\
Improper $S_6$               & 14.7                   & 1                &  & 18.0                       & 1                    \\
2D random 10 points          & 7.2                    & 1.1              &  & 8.9                        & 1                    \\
3D random 50 points          & 16.9                   & 1.2              &  & 14.4                       & 1                    \\ \bottomrule
\end{tabular}

\end{table}

\subsection{Complete Set of Orthogonal Projection Vectors as A Canonical Coordinate Frame}

In 3D, a complete set of orthogonal bases can be found greedily using the protocol as described in Alg.~\ref{alg:orthogonal-projection}.
Specifically, we use the globally optimal solution of the minisum optimization problem as the first basis $\mathbf{b}_\alpha$, and the constrained optimal solution in a plane orthogonal to $\mathbf{b}_\alpha$ as the second basis $\mathbf{b}_\beta$. Special care must be taken for determining the third basis $\mathbf{b}_\gamma$, as the only degree of freedom now is its sign due to the orthogonality constraint. The straightforward approach of choosing the direction that gives the smaller objective function value may fail, for example, when the system contains improper rotational symmetry. In that case, $\mathbf{b}_\alpha$ and $\mathbf{b}_\beta$ are interchangeable and both perpendicular to the rotation-reflection axis. As a result, the two candidates of $\mathbf{b}_\gamma$ will both align with the rotation-reflection axis and are thus indistinguishable by kernel minisum. However, the projection of the atoms into the two seemingly equivalent coordinate frames are not identical, but rather mirror images of each other.
Fortunately, this can be addressed by choosing the direction of the half-space, as created by the plane $\mathbf{b}_\alpha\text{-}\mathbf{b}_\alpha$, that yields the smaller kernel objective function between the \emph{bisector} of $\mathbf{b}_\alpha$ and $\mathbf{b}_\beta$ versus the points lies in that half-space. This rule can also handle general situations with/without symmetry.

\begin{algorithm}[htbp!]
\caption{The procedure for determining a canonical coordinate frame using kernel minisum.}
\label{alg:orthogonal-projection}
\begin{algorithmic}[1]
\Function{GetCanonicalProjection3D}{$\mathbf{E} = [\mathbf{e}_1, \mathbf{e}_2, \ldots, \mathbf{e}_N]$, $\mathbf{G} = [g_1, g_2, \ldots, g_N]$ }
\State $\mathbf{b}_\alpha \gets$ global minimum of \Call{Minisum}{$\mathbf{E}$, $\mathbf{G}$} using multiple runs of Alg.~\ref{alg:MSA} or Alg.~\ref{alg:MEC}
\If {$\mathbf{E}$ contains only 1 point}
  \State construct arbitrary $\mathbf{b}_\beta \perp \mathbf{b}_\alpha$
  \State $\mathbf{b}_\gamma \gets \mathbf{b}_\alpha \times \mathbf{b}_\beta$
\Else
  \State $\mathbf{b}_\beta \gets$ global minimum of \Call{Minisum}{$\mathbf{E}$, $\mathbf{G}$} using multiple runs of Alg.~\ref{alg:MSA} or Alg.~\ref{alg:MEC}, subjecting to the constraint that the probe vector and the gradient are all $\perp \mathbf{b}_\alpha$
  \State $\mathbf{d} \gets ( \mathbf{b}_\alpha + \mathbf{b}_\beta ) / \sqrt{2}$
  \If {$\sum_{\forall \mathbf{e}_i;\ \mathbf{e}_i^\mathsf{T} (\mathbf{b}_\alpha \times \mathbf{b}_\beta) \ge 0} g_i\,\kappa( \mathbf{d}, \mathbf{e}_i ) \le \sum_{\forall \mathbf{e}_i;\ \mathbf{e}_i^\mathsf{T} (\mathbf{b}_\alpha \times \mathbf{b}_\beta) < 0} g_i\,\kappa( \mathbf{d}, \mathbf{e}_i )$}
    \State $\mathbf{b}_\gamma \gets \mathbf{b}_\alpha \times \mathbf{b}_\beta$
  \Else
    \State $\mathbf{b}_\gamma \gets -\mathbf{b}_\alpha \times \mathbf{b}_\beta$
  \EndIf
\EndIf
\State \Return $\mathbf{b}_\alpha$, $\mathbf{b}_\beta$, $\mathbf{b}_\gamma$
\EndFunction
\end{algorithmic}
\end{algorithm}

It is difficult to prove global uniqueness of the kernel minisum solution given the non-convex nature of the exponentiated cosine and square angle kernels.
In fact, it seems that the only kernel that allows analytical proof of solution uniqueness is
$\kappa_\mathrm{COM}(\mathbf{w},\mathbf{e}) \doteq  -\mathbf{w}^\mathsf{T}\mathbf{e}$, whose solution
simply corresponds to the weighted center of mass of the neighbor atoms.
Unfortunately, this simple kernel is not robust against reflectional and rotational symmetry.
Luckily, the rare cases where two global optimal solutions do coexist can be safely captured by the repeated searching procedure starting from different seeds. Thus, a fingerprint can be extracted using each of the resulting coordidate frame. This may mildly increase the size of the training set, which could even be advantagenous when training data is scarce.

\section{Density-Encoded Canonically Aligned Fingerprint}

\label{sec:density-field}

\subsection{Density Field and Approximation of Volume Integral}

\begin{figure}[htbp]
	\centering
	\includegraphics[width=0.9\columnwidth]{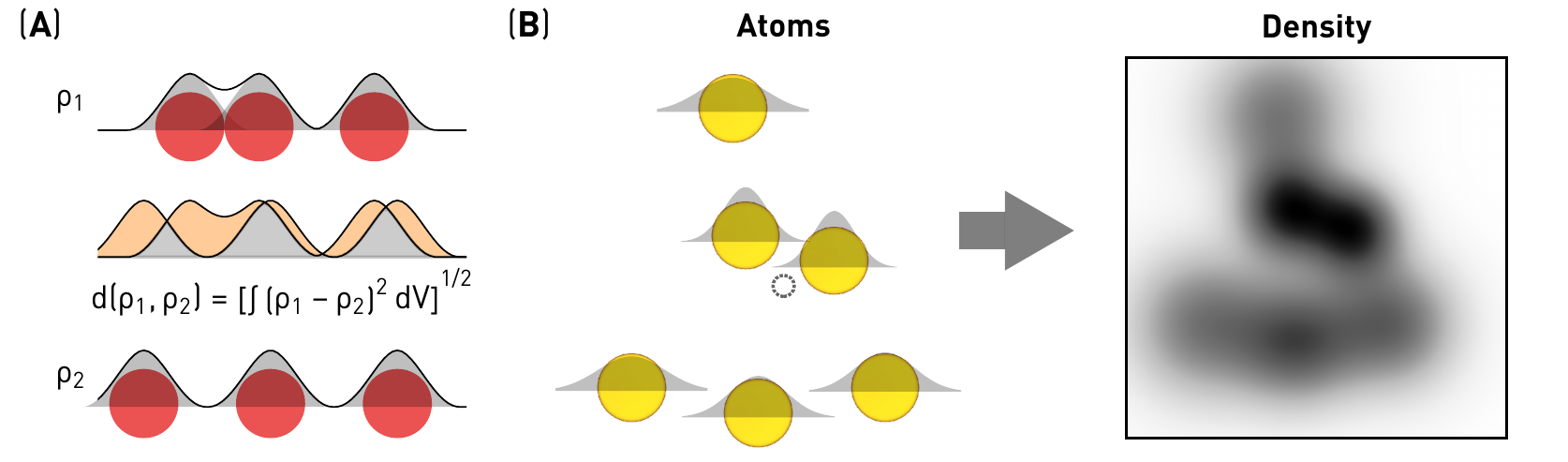}
	\caption{\textbf{(A)} Two 1D density profiles, $\rho_1$ and $\rho_2$, are generated from two different atomistic configurations using atom-centered smoothing kernel functions. The `distance' between them is measured as the $L_2$ norm of their pointwise difference, which corresponds to the orange area in the middle plot. \textbf{(B)} Shown here is a 2D density field using smoothing kernels whose widths depend on the distances of the atoms from the origin. Darker shades indicate higher density.\label{fig:a2d}}
\end{figure}

The local density field $\rho_\mathbf{s}(\mathbf{r})$ around a point $\mathbf{s}$ is formulated as a superimposition of smoothing kernel functions each centered at a neighbor atom $i=1,2,\ldots,N$ with relative displacement $\mathbf{x}_i$ with regard to $\mathbf{s}$ and within a cutoff distance $R_\mathrm{c}$:
\begin{align}
\rho_\mathbf{s}(\mathbf{r}) = \sum_{i,\lVert\mathbf{x}_{i} - \mathbf{s}\rVert < R_\mathrm{c}}^N \Wc(\mathbf{x}_i-\mathbf{s})\,\Wd(\mathbf{x}_i - \mathbf{r})
\end{align}
This density field, as has been pointed out previously~\cite{Bartok2013a}, may be used as a fingerprint of the local atomistic environment.
Here, we assume that the smoothing kernel $\Wd(\mathbf{r})$ takes the form of a stationary Gaussian $\sigma^{-1}\,\exp\left[ -\frac{1}{2} \lVert \mathbf{r} \rVert ^2 / \sigma^2 \right]$. We also assume that the density scaling function $\Wc(\mathbf{r})$, which ensures the continuity of the density field when atoms enter or exit the cutoff distance, is a bell-shaped function with compact support. Further discussion on both $\Wc(\mathbf{r})$ and $\Wd(\mathbf{r})$ can be found in Section~\ref{sec:radial-weight} and Section~\ref{sec:kernel-and-quadrature}, respectively.

To achieve rotational invariance, we project the atom coordinates into the canonical coordinate frame $\mathbf{R} \doteq [\mathbf{b}_\alpha, \mathbf{b}_\beta, \mathbf{b}_\gamma]$ as determined by the kernel minisum algorithm, when generating the density field:
\begin{align}
\boxed{
\rho_\mathbf{s}(\mathbf{r}) = \sum_{i,\lVert\mathbf{x}_{i} - \mathbf{s}\rVert < R_\mathrm{c}}^N \Wc(\mathbf{R}^\mathsf{T}\mathbf{x}_i-\mathbf{s})\, \Wd(\mathbf{R}^\mathsf{T}\mathbf{x}_i - \mathbf{r}). \label{eq:density-field}
}
\end{align}
Depending on the specific application, $\mathbf{s}$ may not necessarily overlap with any of the $\mathbf{x}_i$.
Scalar properties can be acquired directly from the target atom, while vector-valued properties, such as force, can be acquired and interpolated in the local orthogonal coordinates as $\mathbf{\tilde{y}} = \mathbf{R}^\mathsf{T}  \mathbf{y}$.

\newcommand{\drho}[0]{\Delta\rho_{\mathbf{s}_1 \mathbf{s}_2}}
\newcommand{\dRho}[0]{\Delta P_{\mathbf{s}_1 \mathbf{s}_2}}

We define the distance between two density fields $\rho_i$ and $\rho_j$ as a weighted volume integral of their pointwise difference:
\begin{align}
\drho(\mathbf{r}) &\doteq \rho_{\mathbf{s}_1}(\mathbf{r}) - \rho_{\mathbf{s}_2}(\mathbf{r}),\\
\mathrm{d}(\rho_{\mathbf{s}_1},\rho_{\mathbf{s}_2}) &\doteq \left( \int_{\mathbb{R}^3} w(\mathbf{r})\, \left|\drho(\mathbf{r})\right|^2\,\mathrm{d}V\left(\mathbf{r}\right) \right)^{1/2}.\label{eq:integral}
\end{align}
The weight function $w(\mathbf{r})$ provides additional flexibility for emphasizing particular regions of the atomistic neighborhood. It could play an important role for fitting properties with steep gradients, \textit{e.g.} the repulsive part of the Lennard-Jones potential.

\begin{figure}[htbp!]
	\centering
	\includegraphics[width=0.9\columnwidth]{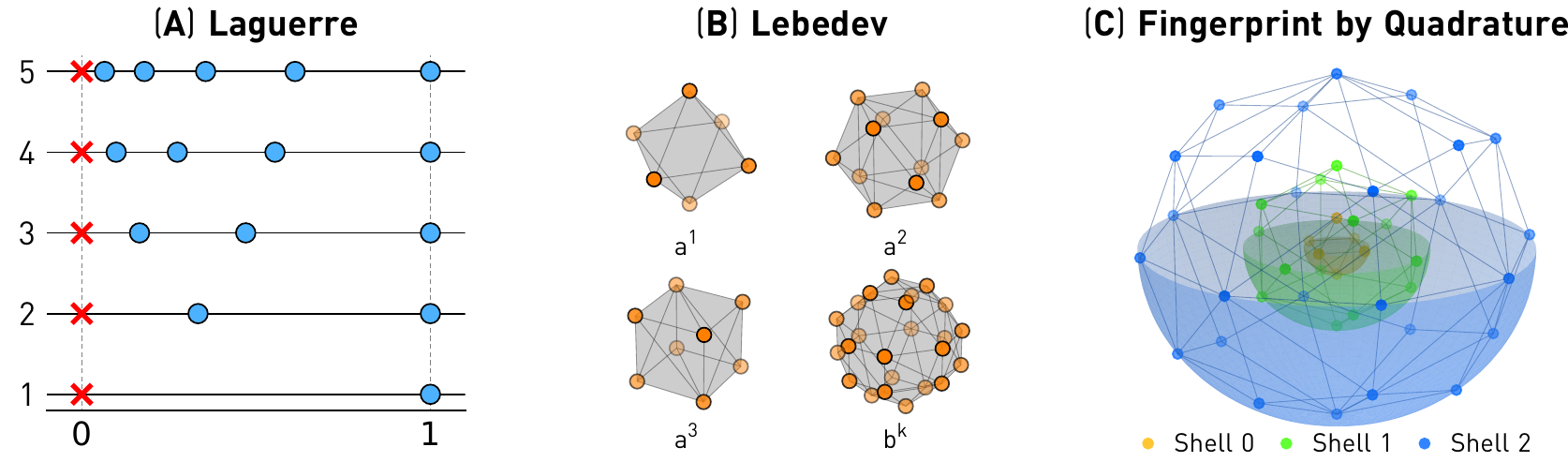}
	\caption{\textbf{(A)} Laguerre quadrature nodes with order 1 to 5, normalized by the reciprocal of the largest node onto the unit interval. \textbf{(B)} The $a^1$, $a^2$, $a^3$, and $b^k$ class of Lebedev grid points on a unit sphere. \textbf{(C)} The DECAF molecular fingerprint essentially comprises of a grid of quadrature nodes that optimally samples the density field induced by the neighbor atoms.
	Shown here is an example of one such composite quadrature grid which combines a 3-node Laguerre quadrature rule with three layers of Lebedev quadrature nodes of 3rd, 5th, and 7th order, respectively. \label{fig:quadrature}}
\end{figure}

We now introduce an optimal quadrature rule to approximate the integral in Eq.~\ref{eq:integral} in a computationally tractable manner. A quadrature rule is a numerical recipe in the form $\int f(x) = \sum_{i=0}^{N} w_i f(x_i)$, which numerically approximates a definite integral using only discrete evaluations of the integrand.
To determine the quadrature nodes and weights, we decompose the volume integral in Eq.~\ref{eq:integral} into a surface integral over spherical shells and a 1D integral along the radial direction:
\begin{align}
\int_{\mathbb{R}^3} w(\mathbf{r})\,\left|\drho(\mathbf{r})\right|^2\,\mathrm{d}V\left(\mathbf{r}\right)
&=
\int_{r=0}^{\infty}
\left(
\int_{\varphi=0}^{2\pi}
\int_{\theta=0}^{\pi}
w(r,\theta,\varphi)\,\left|\drho(r,\theta,\varphi)\right|^2\,
\sin\theta\,
\mathrm{d}\theta\,
\mathrm{d}\varphi\,
\right)
r^2\,
\mathrm{d}r\label{eq:surface-integral}
\end{align}
The surface integral can be optimally approximated using the Lebedev quadrature rule \cite{lebedev1999quadrature}:
\begin{align}
\dRho(r) \doteq \int_{\varphi=0}^{2\pi}
\int_{\theta=0}^{\pi}
w(r,\theta,\varphi)\,\left|\drho(r, \theta,\varphi)\right|^2\,
\sin\theta\,
\mathrm{d}\theta\,
\mathrm{d}\varphi
\approx
4\pi \sum_{m=1}^{N_\mathrm{a}(r)} w(r \cdot \mathbf{q}_m)\,\beta_m \left|\drho(r \cdot \mathbf{q}_m)\right|^2,\label{eq:lebedev}
\end{align}
where $\beta_m$, $\mathbf{q}_m \doteq (x_m, y_m, z_m)$, and $N_\mathrm{a}$ are the weights, positional unit vectors, and number of the Lebedev nodes, respectively.
The radial integral fits well into the generalized Laguerre-Gauss quadrature formula with weight function $r^2\,e^{-r}$ \cite{Rabinowitz}:
\begin{align}
\int_0^\infty \dRho(r)\,r^2\,\mathrm{d}r
\approx
\sum_{n=1}^{N_\mathrm{r}} \alpha_n\,e^{r_n}\,\dRho(r_n), \label{eq:laguerre}
\end{align}
where $\alpha_n$, $r_n$, and $N_\mathrm{r}$ are the weights, coordinates, and number of the Laguerre nodes, respectively.
Combining Eq.~\ref{eq:surface-integral}--\ref{eq:laguerre}, a composite quadrature rule can be generated consisting of several spherical layers of nodes. As shown in Figure~\ref{fig:quadrature}, the radial coordinates of the quadrature nodes are determined by the Laguerre quadrature nodes, while the angular positions are determined by the Lebedev quadrature nodes, respectively. This composite quadrature formula translates the 3D volume integral into a summation over discrete grid points:
\begin{align}
\int_{\mathbb{R}^3} w(\mathbf{r})\,\left|\drho(\mathbf{r})\right|^2\,\mathrm{d}V\left(\mathbf{r}\right)
&\approx 4 \pi \sum_{n=1}^{N_\mathrm{r}} \sum_{m=1}^{N_\mathrm{a}(n)} \alpha_n\,\beta_m\,w( r_n \cdot \mathbf{q}_m )\,e^{r_n}\,\left|\drho(r_n \cdot \mathbf{q}_m)\right|^2. \label{eq:composite}
\end{align}
Using the right hand side of Eq.~\ref{eq:composite} to replace the integral in Eq.~\ref{eq:integral}, and use the multi-index notation $k = (n,m); 1 \le n \le N_\mathrm{r}, 1 \le m \le N_\mathrm{a}(n)$ to enumerate over the quadrature nodes located at $\mathbf{r}_k = r_n \cdot \mathbf{q}_m$ with weights $w_k = 4 \pi\,\alpha_n\,\beta_m\,w(r_n\cdot\mathbf{q}_{m})\,e^{r_n}$, we obtain the final discretized formula for computing the distance between the fingerprints:
\begin{align}
\mathrm{d}(\rho_{\mathbf{s}_1},\rho_{\mathbf{s}_2}) &\approx \left[ \sum_{k=1}^{N} w_{k}\,\left|\rho_{\mathbf{s}_1}(\mathbf{r}_{k}) - \rho_{\mathbf{s}_2}(\mathbf{r}_{k})\right|^2 \right]^{1/2}.
\label{eq:fingerprint-dist}
\end{align}
For quick reference, we tabulated in Appendix the values for $r_n$ and $\alpha_n$ in the Laguerre quadrature of up to 6 points, and the values for $\mathbf{q}_m$, $\beta_m$ in the Lebedev quadrature of up to 50 points.

In addition, the quadrature nodes could be radially scaled such that the outer most nodes lie at a radius $R^*$ within some cutoff distance $R_\mathrm{c}$. This allows us to fit a Laguerre quadrature of any order within an arbitrary cutoff distance.
The scaled quadrature rule is given by:
\begin{align}
\tau &= R^*\                     / \max_n\left({r_\mathrm{n}}\right),\\
\mathrm{d}^*(\rho_{\mathbf{s}_1},\rho_{\mathbf{s}_2}) &\approx \boxed{ \left[ \sum_{k=1}^{N} \tau^3\,w_k\,\left|\rho_{\mathbf{s}_1}(\tau\,\mathbf{r}_k) - \rho_{\mathbf{s}_2}(\tau\,\mathbf{r}_k)\right|^2 \right]^{1/2}. }
\label{eq:fingerprint-dist-scaled}
\end{align}
Since the scaling is simply constant among all nodes, it can be safely omitted in many regression tasks where only the relative distance between the fingerprints are of significance.

\subsection{Radial Weight Functions}
\label{sec:radial-weight}

In this section, we examine two radial weight functions that can be used to fine-tune the density field fingerprint: the density scaling function $\Wc(\mathbf{r})$ as appears in Eq.~\ref{eq:density-field} and the weight of integral $w(r)$ as appears in Eq.~\ref{eq:integral}.

Driven by the interest of reducing computational cost, we would like to use a cutoff distance to select atoms involved in constructing the density field. However, it is important to ensure that atoms will enter and exit the neighborhood smoothly. This naturally requests that the contribution of an atom to be zero outside of the cutoff, and to increase continuously and smoothly when the atom approaching entrance. Correspondingly,  $\Wc(\mathbf{r})$ should: 1) become unity at the origin; 2) smoothly approach zero at the cutoff distance; and 3) be twice-differentiable to ensure the differentiability of regression models based on the fingerprint. Candidates satisfying the above conditions include, for example, a tent-like kernel
\begin{align}
\Wc(\mathbf{r}) = (1 - \lVert \mathbf{r} \rVert / R_\mathrm{c})^t,\ t>2 \label{eq:tent}
\end{align}
and a bell-shaped polynomial kernel with compact support
\begin{align}
\Wc(\mathbf{r}) = \frac{ -b\,(1 - \lVert r \rVert / R_\mathrm{c})^a + a\,(1 - \lVert r \rVert / R_\mathrm{c})^b }{a-b},\ a>b>2
\end{align}
as detailed in Appendix.

The approximation of the radial integral using a Laguerre quadrature requires that the integrand, \textit{i.e.} the pointwise difference between the atomistic density fields, decays sufficiently fast beyond the outermost quadrature nodes in order to achieve acceptable convergence. In addition, the steeply repulsive yet flat attractive interatomic short-range interactions prompt that the sensitivity of fingerprint be adjusted correspondingly in order to avoid numerical difficulties in training a regression model. The weight of the integral, $w(r)$, provides a convenient means for achieving the purpose. Different from $\Wc(\mathbf{r})$, $w(r)$ should instead satisfy the following conditions: 1) is normalized such that $\int w(r) \mathrm{d}V(\mathbf{r}) = 1$; 2) decays sufficiently fast, but not necessarily to 0, beyond the outer most quadrature node; and 3) be sufficiently smooth beyond the outermost quadrature node. Candidates for $w(r)$ includes the tent-like kernel and the bell-shaped kernel for $\Wc(\mathbf{r})$, albeit with a different normalization factor. The Laplacian kernel
\begin{align}
w(r) = \exp(-\left| r \right| / l) / (8\pi l^3),\ l>0
\end{align}
with a properly sized length scale $l$ also appears to be a candidate due to its similarity with $e^{-r}$ part of the weight function of the Laguerre quadrature. Note that the constant kernel
\begin{align}
w(r) = 3 / (4\pi R_\mathrm{c}^3)
\end{align}
may also be a choice as long as the density field already decays fast enough due to the density scaling function $\Wc(\mathbf{r})$.

\begin{figure}[t!]
	\centering
	\includegraphics[width=0.9\columnwidth]{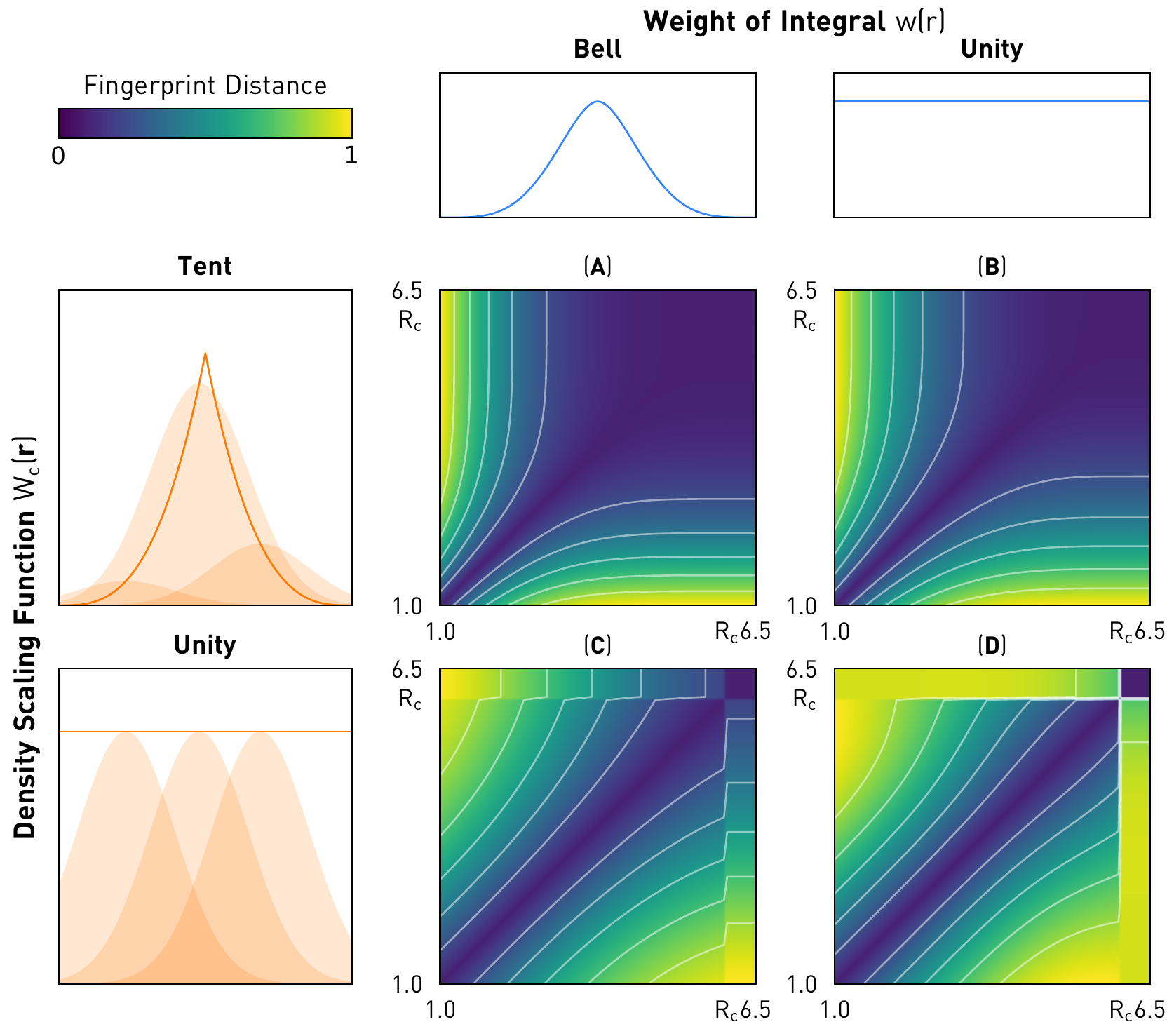}
	\caption{Shown here are examples of the distance matrices between fingerprints sampled from a biatomic system. As manifested by the difference between \textbf{(A)} against \textbf{(B)}, a bell-shaped weight of integral helps to emphasize the near field. Meanwhile, the obvious discontinuities in the second row of matrices demonstrate the importance of the density scaling function when the fingerprint algorithm uses only atoms within a finite cutoff distance.\label{fig:radial-weight}}
\end{figure}

In Figure~\ref{fig:radial-weight}, we demonstrate the effect of the density scaling function and the weight of integral on the distance matrices between fingerprint obtained from a pair of atoms. The comparison between panel A and B shows that a bell-shaped integration weight allows the distance between fingerprints to change more rapidly when the atoms are closer but more slowly when the atoms are farther apart. The visible discontinuity in the second row clearly demonstrates the importance of the damping function when only atoms within a finite cutoff distance are used to compute the fingerprint.

\begin{figure}[htbp!]
	\centering
	\includegraphics[width=0.9\columnwidth]{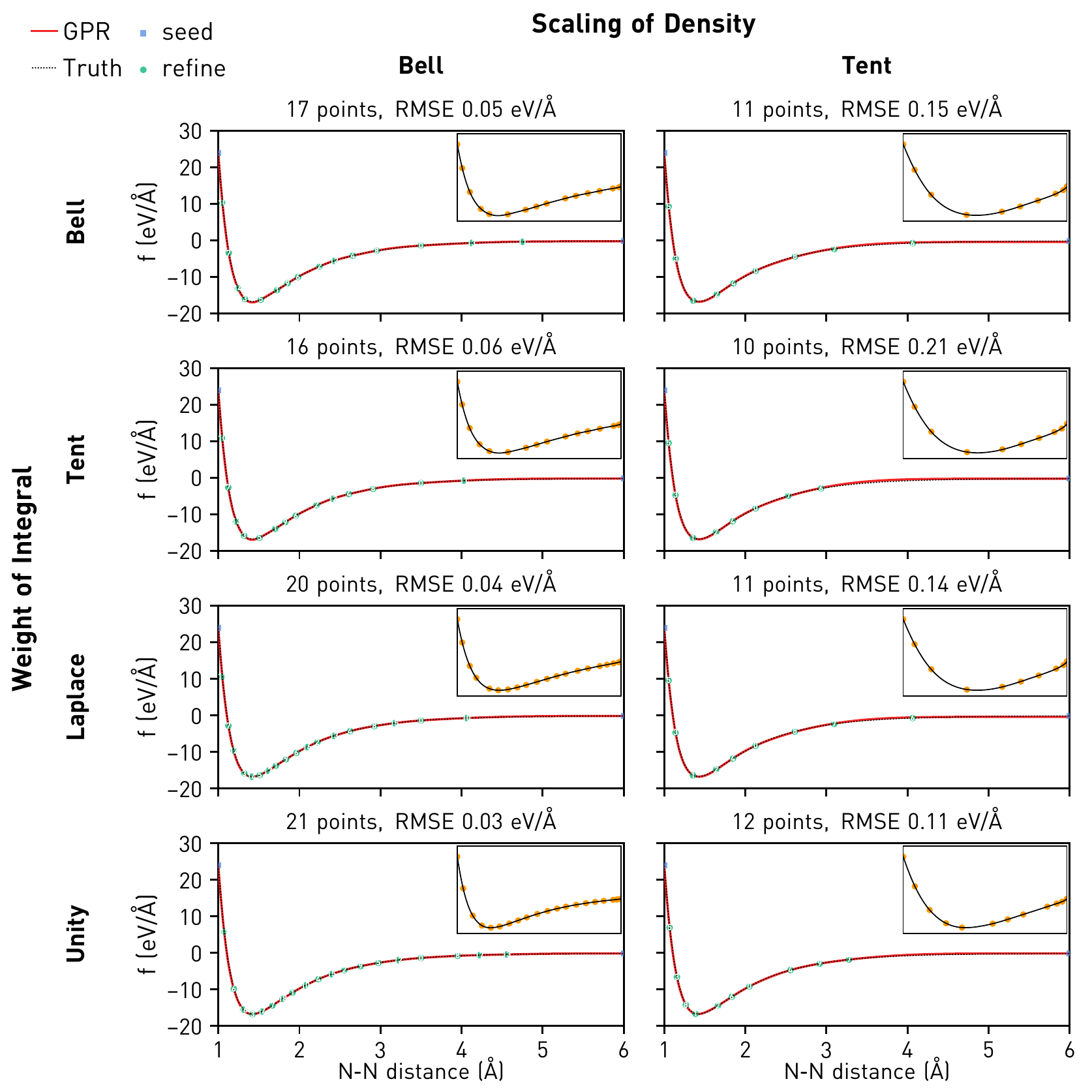}
	\caption{Gaussian process regression of the force between two nitrogen atoms as a function of interatomic distance using different combinations of radial weight functions. Inset figures are plots of the regression function using distances from the feature space. \label{fig:weight-comparison}}
\end{figure}

We further examine the impact of the weight functions on the performance of Gaussian process regression (GPR) using the fingerprint of the interatomic force of a minimal system containing two nitrogen atoms. Despite the simplicity of the system, this case is of fundamental importance because of its ubiquity, and because the fast-growing repulsive regime of the Lennard-Jones potential could cause difficulty as a small change in the system configuration can trigger large changes in the regression target function.
In Figure~\ref{fig:weight-comparison}, we compare the performance among the combination of four weights of the  integral and two density scaling functions. The initial training set consists of two samples collected at $r_\text{N-N} = 1.0$ and $6.0$. The regression is then refined using a greedy strategy that consecutively learns the point with the largest posterior variance until the largest uncertainty, defined as twice the posterior standard deviation, is less than 0.1 $\mathrm{eV}/\r{A}$. The active learning scheme is able to delivery a GPR model, for each and every combination of the weight functions, that closely fits the target function. However, the numbers of refinement iterations and the achieved accuracy do vary. Therefore, it is important to evaluate and choose the weight functions in the context of specific application scenarios.

\subsection{Quadrature Resolution and Density Kernel}%
\label{sec:kernel-and-quadrature}

Despite the formal convergence of the composite quadrature in DECAF, a cost of $\mathcal{O}(NM)$ distance calculations and kernel evaluations are needed to sample a density field generated by $N$ atoms using $M$ quadrature nodes. A less prominent cost is associated with the $L^2$ distance calculation, which comes at a cost of $\mathcal{O}(M)$ floating point operations. Thus, in practice it is often desirable to use as few nodes as possible to capture only information of the density field within a certain band limit\cite{Leistedt2012}. Accordingly, the integral cutoff $R_\mathrm{c}$, the number of quadrature nodes, and the width of the density kernel need to be tuned to obtain an optimal balance between resolution and computational speed.

When designing the composite quadrature rule, we chose the Laguerre quadrature for the radial direction because its nodes are denser near the origin but sparser farther away. This is consistent with our physical intuition that the near field generally has a stronger influence than the far field in an atomistic neighborhood. For example, the Van de Waals potential grows rapidly when atoms are in direct contact, but flattens out of the first coordinate shell. Accordingly, it may be possible for us to use sparser outer-layer grids to reduce the total number of quadrature nodes, while still keeping enough nodes in the inner layers to maintain the sensitivity of the quadrature toward close neighbors. Cooperatively, we can also use non-stationary Gaussian density kernels whose width dependent on the distance from the atom to the origin. In this way, the sparser nodes should still sufficiently sample the smoother far field. Wider kernels at remote atoms also reduce the total difference between the far fields of two fingerprints in a statistical sense. Thus, the contribution of the far field in the integral can be effectively tuned even though the weights on the quadrature nodes remain the same.

\begin{figure}[htbp]
	\centering
	\includegraphics[width=0.9\columnwidth]{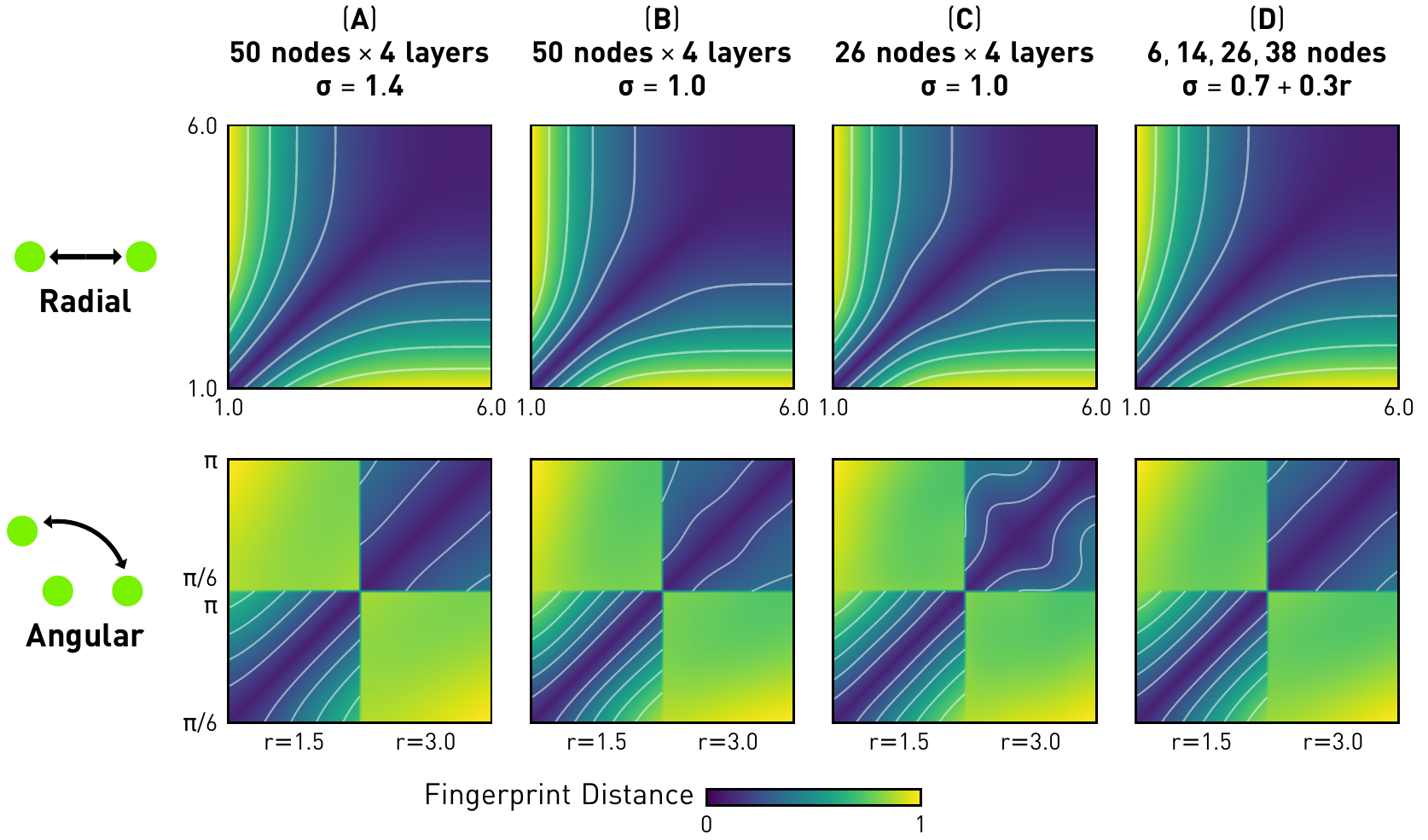}
	\caption{A comparison of fingerprint distance matrices corresponding to bond stretching and angular stretching movements. \textbf{(A)} Dense grid + large smoothing length: radial similarity decreases monotonically while angular similarity changes nearly constantly.
	\textbf{(B)} Dense grid + smaller smoothing length: better fingerprint sensitivity in the near field for bond stretching, compromised linearity in the far field for angular stretching. \textbf{(C)} Sparser grid + smaller smoothing length: compromised far-field performance for both bond and angular movements. \textbf{(D)} Variable-resolution grid + radially dependent smoothing  length: good resolution and linearity in both near and far fields.\label{fig:quadrature-resolution}}
\end{figure}

In Figure~\ref{fig:quadrature-resolution}, we demonstrate how a variable-resolution quadrature can be combined with a widening smoothing density kernel to simultaneously reduce the computational complexity and preserve the quality of the fingerprint.
In column A, a dense grid is used to sample density fields generated by a wide smoothing length. By examining the distance matrices of fingerprints sampled during bond stretching and angular stretching movements, we note that the radial similarity decreases monotonically while the angular similarity changes nearly constantly.
In column B, the number of quadrature nodes is kept the same, but the smoothing length is reduced as an attempt to increase fingerprint sensitivity. Better response in the near field of the radial direction is obtained, but the linearity in the far field in the angular direction is compromised.
In column C, the fingerprint performs even worse due to the combination of a sparser quadrature grid and a small smoothing length. In column D, the performance recovered because we let the smoothing length parameter $\sigma$ of the Gaussian density kernels $\Wd(\mathbf{r})$ depend on the distance from the origin to each atom, and simultaneously adjust the quadrature node density according to this pattern.

\FloatBarrier

\section{Demonstration}
\label{sec:demonstration}

\subsection{Method}

\begin{figure}[htbp!]
	\centering
	\includegraphics[width=0.9\columnwidth]{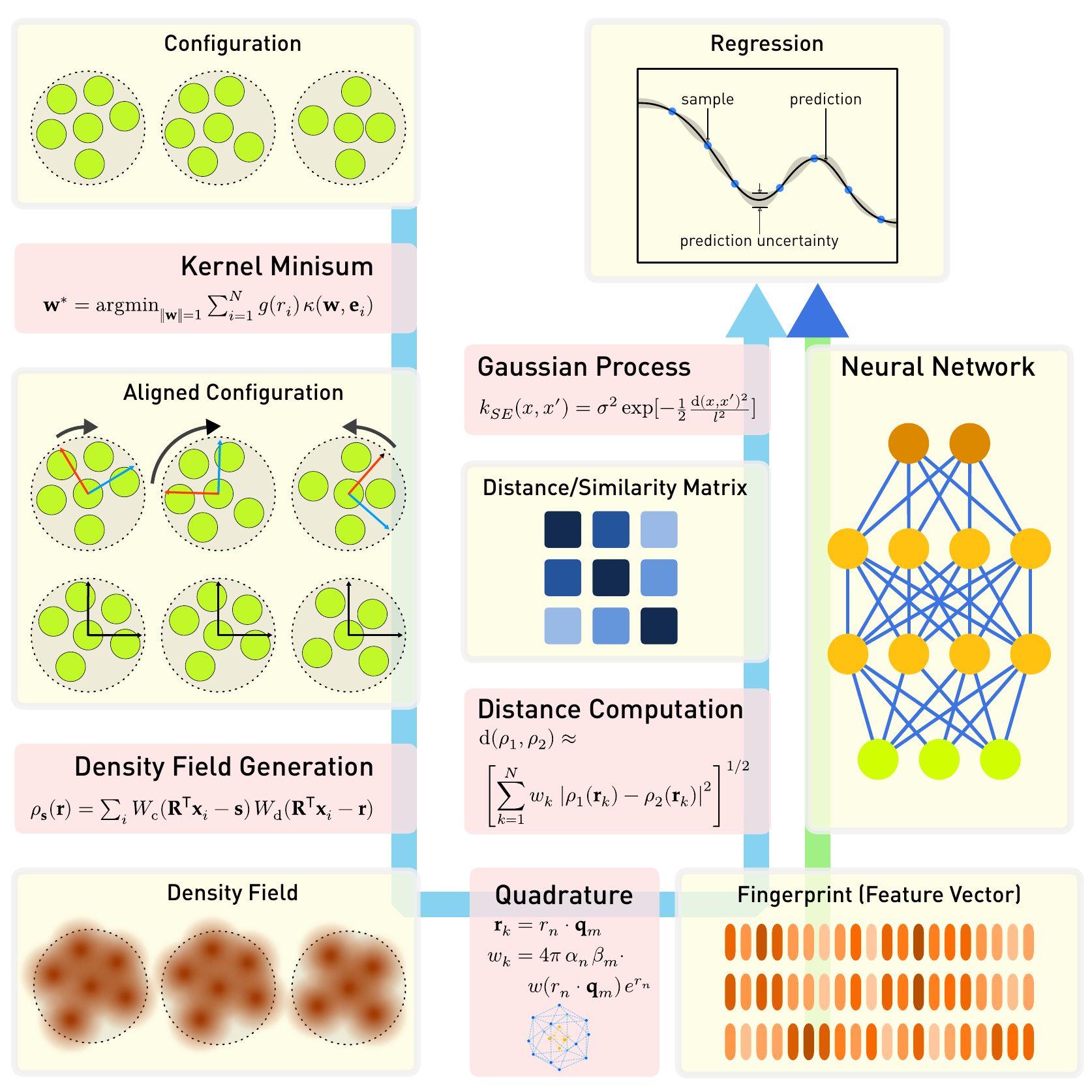}
	\caption{Shown here is the workflow of regression using the density field fingerprint. The key stages, \textit{i.e.} kernel minisum optimization, density field generation, and quadrature-based distance computation, are covered in detail in Sec.~\ref{sec:cframe} and Sec.~\ref{sec:density-field}. The fingerprints can also be readily used to train artificial neural networks.\label{fig:workflow}}
\end{figure}

Regression tasks throughout this work are performed using Gaussian process regression (GPR), a nonlinear kernel method that treats training data points as discrete observations from an instantiation of a Gaussian process. Predictions are made using the posterior mean and variance of the joint distribution between the test data and the training data. One particular interesting property about Gaussian process is that the posterior variance may be interpreted as a measure of prediction uncertainty, which can be exploited to design active learning algorithms for sampling expensive functions. The actual computation used our own software implementation which was made publicly available on Zenodo~\cite{Tang2017a}. We use the square exponential covariance kernel to compute the covariance, \textit{i.e.} correlation, between the samples:
\begin{align*}
k_{SE}(x,x') = \sigma^2 \exp[ -\frac{1}{2} \mathrm{d}(x, x')^2 / l^2 ]
\end{align*}
where $x$ and $x'$ are DECAF fingerprints, and $\mathrm{d}(x, x')$ the distance between norm as computed by Eq.~\ref{eq:fingerprint-dist} or Eq.~\ref{eq:fingerprint-dist-scaled}.
The kernel is stationary, meaning that the covariance depends only on the relative distance between two samples but not their absolute position. The training process searches for the hyperparameters, \textit{i.e.} the output variance $\sigma$ and the length scale $l$, that maximizes the likelihood of the training data. A detailed tutorial on GPR can be found in Ref.~\cite{Rasmussen2006a}. An illustration on the complete workflow of using the density field fingerprint to perform regression tasks is given in Figure~\ref{fig:workflow}.

\subsection{Potential Energy Surface}

\begin{figure}[htbp!]
	\centering
	\includegraphics[width=0.9\columnwidth]{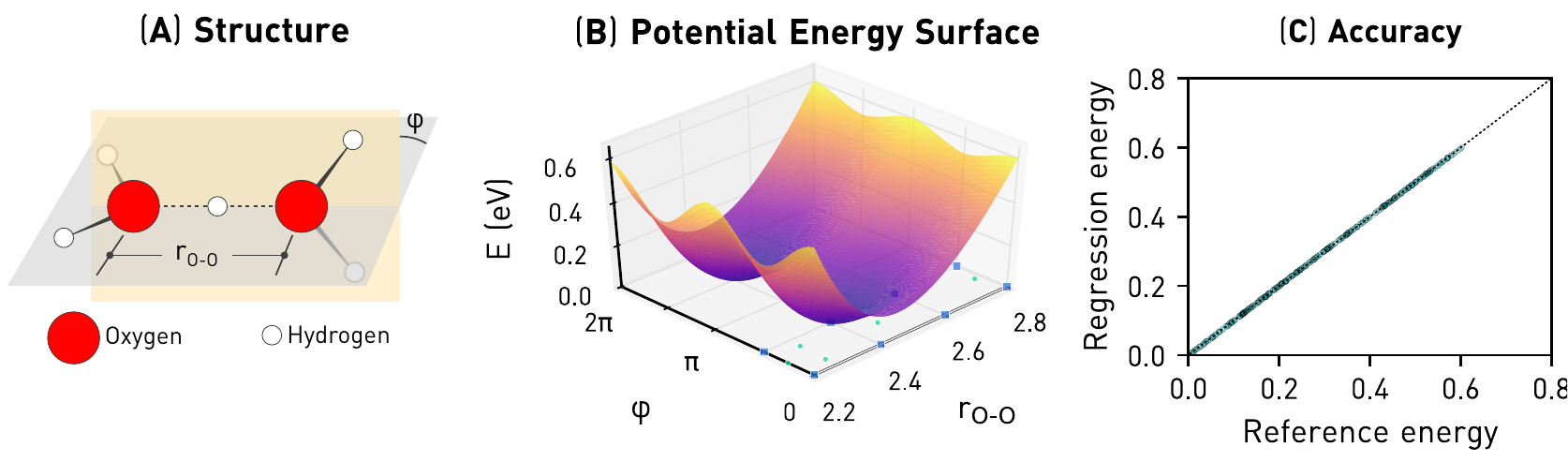}
	\caption{Gaussian process regression is carried out to fit the potential energy surface of a protonated water dimer as a function of two internal variables, \textit{i.e.} the oxygen-oxygen distance $r_\mathrm{O\text{-}O}$ and the dihedral angle $\varphi$ between the plane determined by the two water molecules. This system contains an improperly rotational symmetry, which can be correctly recognized by the kernel minisum-based algorithm given in Alg.~\ref{alg:orthogonal-projection}. Only a quarter of the domain was used to train the GP, yet the model can accurately predict the energy of the entire parameter space thanks to symmetry detection. \label{fig:validation-pwdimer}}
\end{figure}

First, we attempt to fit the potential energy surface of a protonated water dimer system, in a head-to-head configuration, as a function of the oxygen-oxygen distance $r_{\mathrm{O}\text{-}\mathrm{O}}$ and the dihedral angle $\varphi$ between the two planes each formed by a water molecule. As shown in Figure~\ref{fig:validation-pwdimer}A, the system contains an improperly rotational symmetry, which we wish to capture with the kernel minisum algorithm. A GPR model was seeded with 8 training points corresponding to the combinations of $r_{\mathrm{O}\text{-}\mathrm{O}}=2.2, 2.4, 2.6, 2.8$ and $\varphi=0,\pi/2$. Subsequently, an active learning protocol was used to greedily absorb points with the highest uncertainty into the training set. Despite that we restricted all training data to be within the subdomain $\varphi <= \pi/2$, as shown by Figure~\ref{fig:validation-pwdimer}B and \ref{fig:validation-pwdimer}C, we are able to accurately reproduce the target function over the entire parameter space after a few active learning steps.

The DECAF fingerprint used here is constructed with 3 spherical layers within a cutoff distance $R_c$ of 6.0 \r{A}, each consisting of 14, 26, and 38 Lebedev quadrature nodes, respectively.
The weight of integral was chosen as $w(r) = W^{6,4}( 1 - r / R_c )$, where $W^{6,4}$ is the bell-shaped polynomial as defined in Appendix Eq.~\ref{eq:poly-kernel}. The density scaling function $\Wc(\mathbf{r}) = (1 - \lVert \mathbf{r} \rVert / R_\mathrm{c})^3$, where $\mathbf{r}$ is the vector from the atom to the fingerprint center, is the tent-like kernel as defined in Eq.~\ref{eq:tent} with $t = 3$.
The density kernel that sits on the oxygen atoms assumes the form of a non-stationary Gaussian as discussed in Section~\ref{sec:kernel-and-quadrature}: $\Wd^{O}(\mathbf{r}, \mathbf{r}') = \sigma_{O}(\mathbf{r})^{-1}\,\exp\left[ -\frac{1}{2} \lVert \mathbf{r}' \rVert ^2 / \sigma_{O}(\mathbf{r})^2 \right],\ \sigma_{O}(\mathbf{r}) = 1.5 + 0.25 \lVert \mathbf{r} \rVert$ with $\mathbf{r}$ being the vector from the atom to the fingerprint center and $\mathbf{r}'$ being the vector from the atom to the quadrature node.
The density kernel for the hydrogen atoms has a different weight and width to ensure discriminability: $\Wd^{H}(\mathbf{r}, \mathbf{r}') = 0.75\, \sigma_{H}(\mathbf{r})^{-1}\,\exp\left[ -\frac{1}{2} \lVert \mathbf{r}' \rVert ^2 / \sigma_{H}(\mathbf{r})^2 \right],\ \sigma_{H}(\mathbf{r}) = 0.9 + 0.15 \lVert \mathbf{r} \rVert$.

\subsection{Geometry Optimization and Vibrational Analysis}

Next, we demonstrate the usability of fingerprint for fitting vector-valued quantities by performing geometry optimization and vibrational analysis on a single water molecule. The process involves the simultaneous regression of: 1) energy, a molecular scalar quantity; 2) force, a per-atom vector quantity; and 3) dipole, a molecular vector quantity. Correspondingly, we performed GPR of energy and dipole using fingerprints extracted from the center of mass of the molecule, and GPR of force using fingerprints extracted from each atom. Each component of the vector properties is modeled independently as a scalar Gaussian process. The training set consists of 45 configurations uniformly covering the range $r_\mathrm{O\text{-}H} = 0.93, 0.95, 0.97, 0.99, 1.05 \r{A}$ and $\theta_\mathrm{H\text{-}O\text{-}H} = 101\degree, 105.5\degree, 111\degree$. As shown in Table~\ref{table:water-ir}, the GPR model can successful drive calculations of the infrared spectrum of the molecule from randomly perturbed initial structures in arbitrary orientation. The fingerprint configuration is the same as in the previous section.

\begin{table}[]
\centering
\caption{Geometry optimization and vibrational analysis of a single water molecule using GPR and our proposed fingerprint algorithm. 256 independent trials were performed using coordinates of water perturbed from equilibrium by a Gaussian noise $\mathcal{N}(\mathbf{0},0.15\mathbf{I})$ followed by a randomly chosen rigid-body rotation.}
\label{table:water-ir}
\begin{tabular}{@{}llllll@{}}
\toprule
\multicolumn{2}{l}{}                  & \multicolumn{2}{l}{GPR}                            & DFT \\
\midrule
\multicolumn{2}{l}{Zero-point energy} & \multicolumn{2}{l}{0.591 $\pm$ 0.003 eV}           & 0.583 eV \\
\multicolumn{2}{l}{Static dipole}     & \multicolumn{2}{l}{2.1580 $\pm$ 0.0001 D}          & 2.159 D \\
\multicolumn{2}{l}{Residual Force}    & \multicolumn{2}{l}{0.0016 $\pm$ 0.0005 eV/$\r{A}$} & 0.0003 eV/$\r{A}$ \\
\midrule
\multirow{2}{*}{Mode} & \multicolumn{2}{l}{Frequency (cm\textasciicircum -1)} &  & \multicolumn{2}{l}{Intensity (D/A)\textasciicircum 2 amu\textasciicircum -1} \\ \cmidrule(l){2-6}
                      & GPR                           & DFT                   &  & GPR                                        & DFT                             \\ \midrule
0                     & 1576.5 $\pm$ 1.4              & 1602.4                &  & 1.5726 $\pm$ 0.0005                        & 1.5767                          \\
1                     & 3819.3 $\pm$ 0.9              & 3817.5                &  & 0.2516 $\pm$ 0.0005                        & 0.2159                          \\
2                     & 3916.7 $\pm$ 1.6              & 3922.6                &  & 1.3349 $\pm$ 0.0028                        & 1.3401                          \\ \bottomrule
\end{tabular}
\end{table}

\subsection{Molecular Dyanmics Trajectory}

\begin{figure}[htbp!]
	\centering
	\includegraphics[width=0.9\columnwidth]{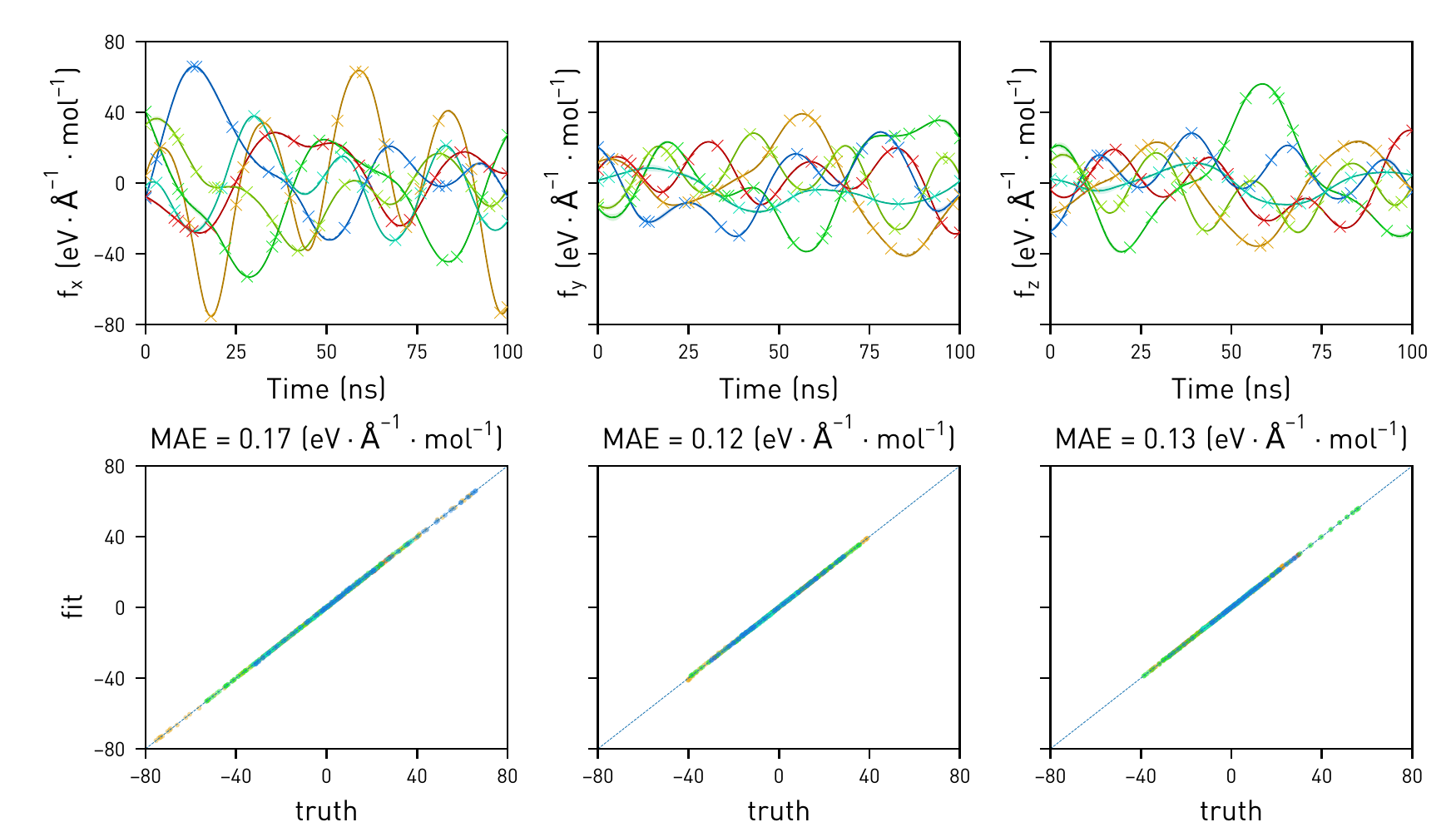}
	\caption{The force exerted on the carbon atoms in a benzene molecule was fit using 100 out of the 1200 configurations in a 100-ns molecular dynamics trajectory at 0.5 ns time step size. Color coding corresponds to carbon atom id, while crosses indicate the atomistic configurations used in the GPR training.\label{fig:validation-force}}
\end{figure}

As shown in Figure~\ref{fig:validation-force}, here we attempt to fit for the forces felt by the atoms in a benzene molecule along the MD Trajectories as obtained from a sibling database of QM7~\cite{Schutt2017,Chmiela2017}. The density kernel for the carbon atoms assumes the same functional form with that of the oxygen atoms, but uses a different smoothing length function $\sigma_{C}(\mathbf{r}) = 1.2 + 0.2 \lVert \mathbf{r} \rVert$. The rest of the parameters are inherited from the previous examples. The training configurations were chosen adaptively in an iterative process using the sum of the GPR posterior variance and the prediction error as the acquisition function.

\section{Connection to Other Fingerprint Algorithms}
\label{sec:discussion}

\begin{figure}[htb!]
	\centering
	\includegraphics[width=0.9\columnwidth]{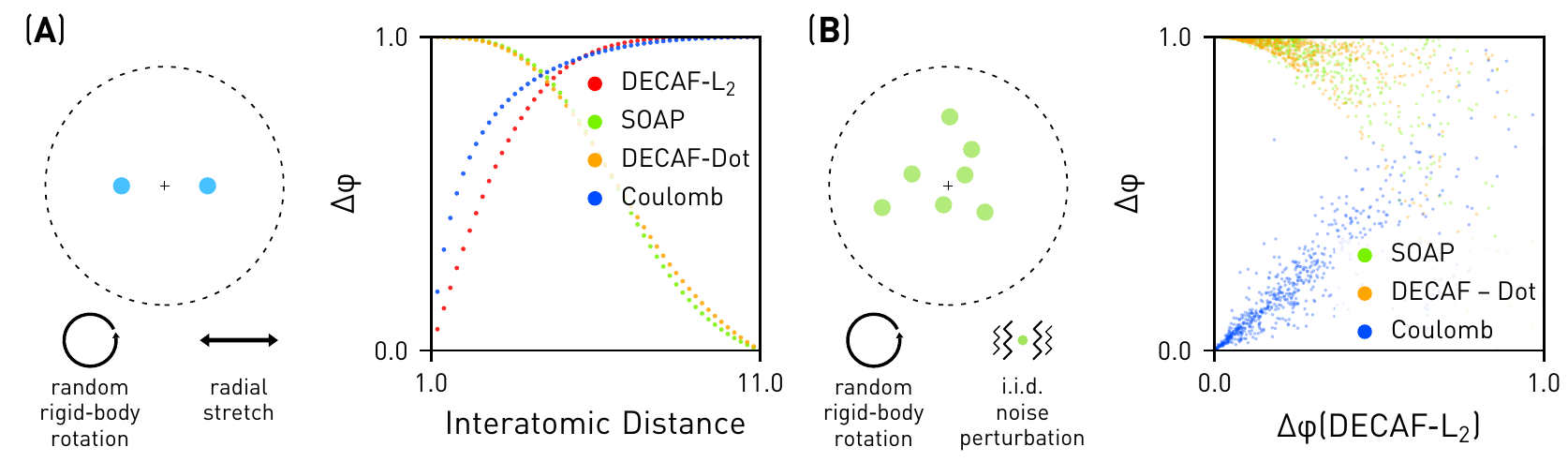}
	\caption{A comparison between the distance measure used DECAF, SOAP, and the Coulomb matrix. Fingerprint distances $\Delta\varphi$ shown in the plots are measured against $\mathbf{r}_{ij} = 1.0$ in \textbf{(A)} and an randomly chosen initial state in \textbf{(B)}. \label{fig:cafe-vs-others}}
\end{figure}

In Figure~\ref{fig:cafe-vs-others}, we compare the ability to distinguish atomistic configurations of our fingerprint as well as SOAP and the Coulomb matrix.
Our work is inspired by the SOAP descriptor \cite{Bartok2013a}, which proposes the use of smoothed densities to represent atomistic neighborhoods. However, instead of converting the density field into the frequency domain using spherical harmonics, we perform density field sampling and comparison directly in the real space. This is enabled thanks to the available of canonical coordinate frame as computed through the kernel minisum optimization. We have mainly used the $L_2$ norm to compute the distance between atomistic neighborhoods. However, our fingerprint exhibits very similar behavior to SOAP when used together with an inner product formula
\begin{align}
\mathrm{d}(\rho_{\mathbf{s}_1},\rho_{\mathbf{s}_2}) &\approx \frac{ \sum_{k=1}^{N} w_{k}\,\rho_{\mathbf{s}_1}(\mathbf{r}_{k})\,\rho_{\mathbf{s}_2}(\mathbf{r}_{k}) } { \sqrt{\sum_{k=1}^{N} w_{k}\,\rho_{\mathbf{s}_1}(\mathbf{r}_{k})\,\rho_{\mathbf{s}_1}(\mathbf{r}_{k})}  \sqrt{\sum_{k=1}^{N} w_{k}\,\rho_{\mathbf{s}_2}(\mathbf{r}_{k})\,\rho_{\mathbf{s}_2}(\mathbf{r}_{k})} }
\end{align}
as demonstrated in Figure~\ref{fig:cafe-vs-others}A. Thus, our fingerprint could be used in conjunction with a wide variety of covariance functions based on either the Euclidean distance or the inner product similarity.

At first sight, DECAF is very different from the Coulomb matrix fingerprint and GRAPE, which are both graph-based algorithms~\cite{Rupp2012,Ferre2017}.
However, instead of trying to capture the overall density field, if we measure the contribution from each individual atom on the quadrature nodes at $\mathbf{z}_1,\mathbf{z}_2,\ldots,\mathbf{z}_{M}$ as a row vector, and stacked up the results to yield the matrix
\begin{align}
\mathbf{E}_{ij}^{N\times M} = \ \mathrm{k}(\mathbf{x}_i, \mathbf{z}_j).
\end{align}
Then $\mathbf{E}$ can be regarded as an incidence matrix \cite{Gross2005} between atoms and the quadrature nodes. This is similar to the graph-based abstraction as seen in the Coulomb matrix and the GRAPE kernel. However, in both cases the vertices in the graph represent atoms while the edges represent pairwise interatomic interactions. Here, the density-based incidence matrix adopts the opposite pattern and constructs a graph with the quadrature nodes being vertices and atoms being edges. The adjacency matrix in this case is computed as the inner product $\mathbf{E}^\mathsf{T}\mathbf{E}$:
\begin{align}
\mathbf{A}_{ij}^{M\times M} = (\mathbf{E}^\mathsf{T}\mathbf{E})_{ij} = \sum_{k=1}^{N} \mathrm{k}(\mathbf{x}_k, \mathbf{z}_i)\,\mathrm{k}(\mathbf{x}_k, \mathbf{z}_j).
\end{align}
The weight on the edges, as represented by the elements of the adjacency matrix $A$, can be interpreted as the total flux as contributed by all paths each bridged by an atom $k$.
We have numerically found that the smallest $N$ eigenvalues (except for the 0 eigenvalue) of the symmetric normalized Laplacian
\begin{align}
\mathbf{L} = \mathbf{I} - \mathbf{D}^{-1/2}\mathbf{A}\mathbf{D}^{-1/2},\ \text{where}\ \mathbf{D}_{ii} = \delta_{ij} \sum_j \mathbf{A}_{ij}
\end{align}
is invariant under rotation up to a certain noise level, even if the quadrature nodes do not rotate with the atoms. Nonetheless, this detour appears to represent a pure theoretical interest rather than any practical value.

\begin{figure}[t!]
	\centering
	\includegraphics[width=0.9\columnwidth]{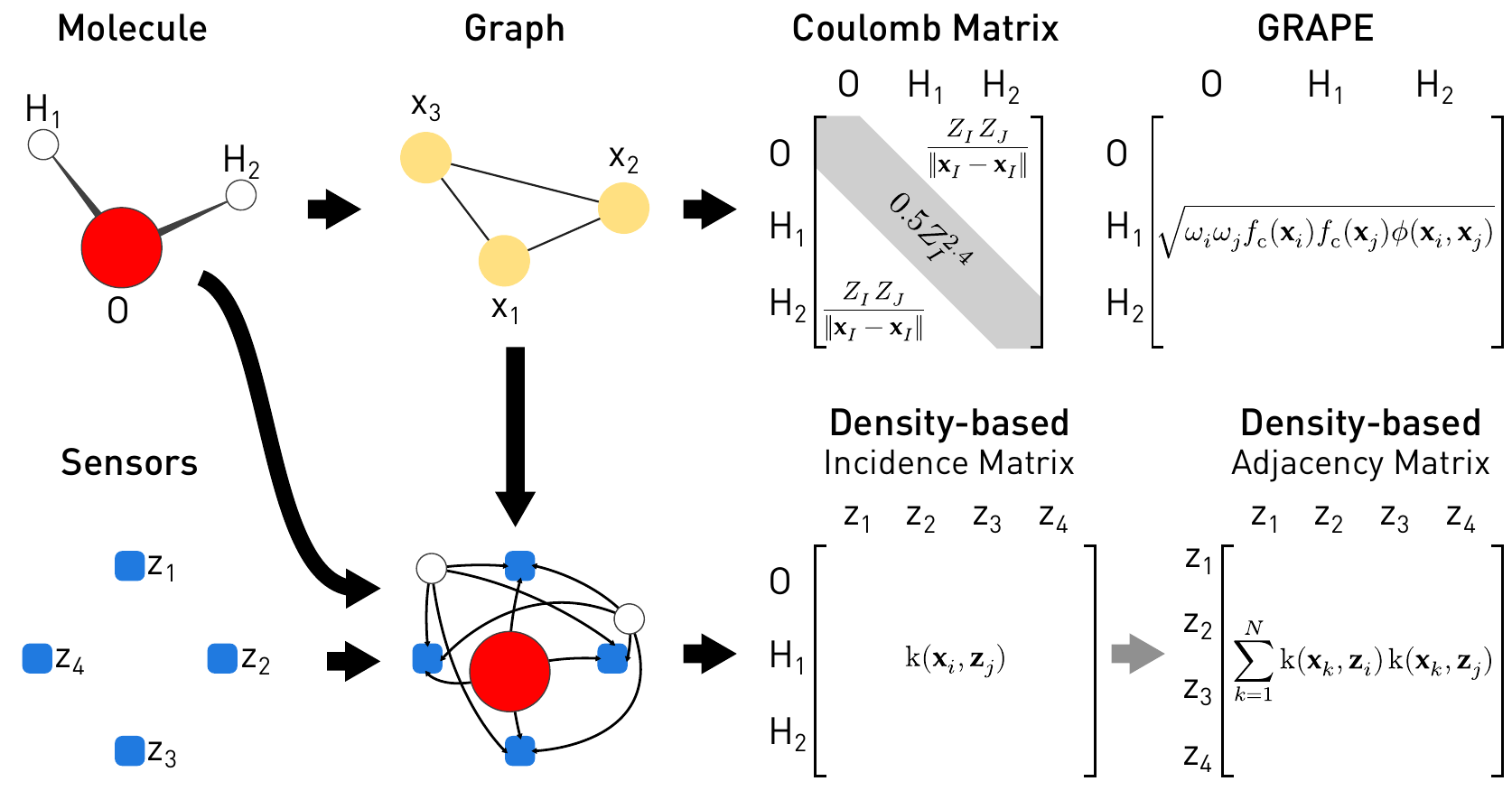}
	\caption{A comparison between graph-based molecular fingerprints. The Coulomb Matrix and the GRAPE kernel construct graphs where nodes corresponds to atoms while the weights on the edges are determined by some pairwise inter-atomic interactions. In contrast, in the density-based incidence matrix we construct a graph on a set of quadrature nodes whose connectivity is weighted by a sum of contributions from individual atoms.\label{fig:matrices}}
\end{figure}

\section{Conclusion}
\label{sec:conclusion}

In this paper, we presented the Density-Encoded Canonically Aligned Fingerprint (DECAF) by exploring the idea of using smoothed density fields to represent and compare atomistic neighborhoods. One of the key enabling technique in DECAF is a kernel minisum algorithm, which allows the unambiguous identification of a canonically aligned coordinate frame that can be used for rotationally invariant projection of the density field as well as any associated vector quantities. We have performed detailed analysis to study the behavior of the fingerprint by changing various parameter, such as resolution, smoothing length, and the choice of weight functions. We demonstrate that the fingerprint algorithm can be used to implement highly accurate regressions of both scalar and vector properties of atomistic systems including energy, force and dipole moment, and could be a useful building block for constructing data-driven next generation force fields to accelerate molecular mechanics calculations with an accuracy comparable to those driven by quantum mechanical theories and calculators.

\section*{Acknowledgment}
\noindent This work was supported by the Department of Energy (DOE) Collaboratory on Mathematics for Mesoscopic Modeling of Materials (CM4). This work was also supported by the Army Research Laboratory under Cooperative Agreement Number W911NF-12-2-0023.

\bibliography{../lit/library}

\begin{thebibliography}{10}

\bibitem{zhao2013mature}
G.~Zhao, J.~R. Perilla, E.~L. Yufenyuy, X.~Meng, B.~Chen, J.~Ning, J.~Ahn,
  A.~M. Gronenborn, K.~Schulten, C.~Aiken, and Others.
\newblock {Mature HIV-1 capsid structure by cryo-electron microscopy and
  all-atom molecular dynamics}.
\newblock {\em Nature}, 497(7451):643--646, 2013.

\bibitem{Lindorff-Larsen2016}
K.~Lindorff-Larsen, P.~Maragakis, S.~Piana, and D.~E. Shaw.
\newblock {Picosecond to Millisecond Structural Dynamics in Human Ubiquitin}.
\newblock {\em The Journal of Physical Chemistry B}, 120(33):8313--8320, 2016.

\bibitem{Rappe1992a}
A.~K. Rappe, C.~J. Casewit, K.~S. Colwell, W.~A. Goddard, and W.~M. Skiff.
\newblock {UFF, a full periodic table force field for molecular mechanics and
  molecular dynamics simulations}.
\newblock {\em Journal of the American Chemical Society}, 114(25):10024--10035,
  1992.

\bibitem{Cornell1995}
W.~D. Cornell, P.~Cieplak, C.~I. Bayly, I.~R. Gould, K.~M. Merz, D.~M.
  Ferguson, D.~C. Spellmeyer, T.~Fox, J.~W. Caldwell, and P.~A. Kollman.
\newblock {A second generation force field for the simulation of proteins,
  nucleic acids, and organic molecules}.
\newblock {\em Journal of the American Chemical Society}, 117(19):5179--5197,
  1995.

\bibitem{WilliamL.Jorgensen1996}
W.~L. Jorgensen, D.~S. Maxwell, and J.~Tirado-Rives.
\newblock {Development and Testing of the OPLS All-Atom Force Field on
  Conformational Energetics and Properties of Organic Liquids}.
\newblock {\em Journal of the American Chemical Society}, 118(15):11225--11236,
  1996.

\bibitem{frenkel2001understanding}
D.~Frenkel and B.~Smit.
\newblock {\em {Understanding molecular simulation: from algorithms to
  applications}}, volume~1.
\newblock Academic press, 2001.

\bibitem{leach2001molecular}
A.~R. Leach.
\newblock {\em {Molecular modelling: principles and applications}}.
\newblock Pearson education, 2001.

\bibitem{Cheng2014}
T.~Cheng, A.~Jaramillo-Botero, W.~A. Goddard, and H.~Sun.
\newblock {Adaptive accelerated ReaxFF reactive dynamics with validation from
  simulating hydrogen combustion}.
\newblock {\em Journal of the American Chemical Society}, 136(26):9434--9442,
  2014.

\bibitem{Braun2014}
D.~Braun, S.~Boresch, and O.~Steinhauser.
\newblock {Transport and dielectric properties of water and the influence of
  coarse-graining: Comparing BMW, SPC/E, and TIP3P models}.
\newblock {\em The Journal of Chemical Physics}, 140(6):064107, 2014.

\bibitem{Boonstra2016}
S.~Boonstra, P.~R. Onck, and E.~van~der Giessen.
\newblock {CHARMM TIP3P Water Model Suppresses Peptide Folding by Solvating the
  Unfolded State}.
\newblock {\em The Journal of Physical Chemistry B}, 120(15):3692--3698, 2016.

\bibitem{Behler2011}
J.~Behler.
\newblock {Atom-centered symmetry functions for constructing high-dimensional
  neural network potentials}.
\newblock {\em The Journal of Chemical Physics}, 134(134), 2011.

\bibitem{Bartok2013a}
A.~P. Bart{\'{o}}k, R.~Kondor, and G.~Cs{\'{a}}nyi.
\newblock {On representing chemical environments}.
\newblock {\em Physical Review B}, 87(18):184115, 2013.

\bibitem{Li2015c}
Z.~Li, J.~R. Kermode, and A.~{De Vita}.
\newblock {Molecular Dynamics with On-the-Fly Machine Learning of
  Quantum-Mechanical Forces}.
\newblock {\em Physical Review Letters}, 114(9):096405, 2015.

\bibitem{Khorshidi2016}
A.~Khorshidi and A.~A. Peterson.
\newblock {Amp: A modular approach to machine learning in atomistic
  simulations}.
\newblock {\em Computer Physics Communications}, 207:310--324, 2016.

\bibitem{Rasmussen2006a}
C.~E. Rasmussen and C.~K.~I. Williams.
\newblock {Gaussian Processes for Machine Learning}.
\newblock 2006.

\bibitem{Specht1991}
D.~Specht.
\newblock {A general regression neural network}.
\newblock {\em IEEE Transactions on Neural Networks}, 2(6):568--576, 1991.

\bibitem{Rupp2012}
M.~Rupp, A.~Tkatchenko, K.-R. Muller, O.~A. von Lilienfeld, K.-R.~R.
  M{\"{u}}ller, O.~{Anatole Von Lilienfeld}, and O.~A. von Lilienfeld.
\newblock {Fast and accurate modeling of molecular atomization energies with
  machine learning.}
\newblock {\em Physical Review Letters}, 108(5):58301, 2012.

\bibitem{Ferre2017}
G.~Ferr{\'{e}}, T.~Haut, and K.~Barros.
\newblock {Learning molecular energies using localized graph kernels}.
\newblock {\em The Journal of Chemical Physics}, 146(11):114107, 2017.

\bibitem{Sun2014}
H.~Y. Sun.
\newblock {\em {Learning over Molecules : Representations and Kernels}}.
\newblock PhD thesis, Harvard University, 2014.

\bibitem{Coifman2005}
R.~R. Coifman, S.~Lafon, A.~B. Lee, M.~Maggioni, B.~Nadler, F.~Warner, and
  S.~W. Zucker.
\newblock {Geometric diffusions as a tool for harmonic analysis and structure
  definition of data: diffusion maps.}
\newblock {\em Proceedings of the National Academy of Sciences of the United
  States of America}, 102(21):7426--31, 2005.

\bibitem{Bartok2010}
A.~P. Bart{\'{o}}k, M.~C. Payne, R.~Kondor, and G.~Cs{\'{a}}nyi.
\newblock {Gaussian Approximation Potentials: The Accuracy of Quantum
  Mechanics, without the Electrons}.
\newblock {\em Physical Review Letters}, 104(13):136403, 2010.

\bibitem{Botu2015}
V.~Botu and R.~Ramprasad.
\newblock {Learning scheme to predict atomic forces and accelerate materials
  simulations}.
\newblock {\em Physical Review B}, 92(9):094306, 2015.

\bibitem{BARZILAI1988}
J.~Barzilai and J.~M. Borwein.
\newblock {Two-Point Step Size Gradient Methods}.
\newblock {\em IMA Journal of Numerical Analysis}, 8(1):141--148, 1988.

\bibitem{lebedev1999quadrature}
D.~{Lebedev, VI and Laikov}.
\newblock {A quadrature formula for the sphere of the 131st algebraic order of
  accuracy}.
\newblock {\em Doklady. Mathematics}, 59(3):477--481, 1999.

\bibitem{Rabinowitz}
P.~Rabinowitz and G.~Weiss.
\newblock {Tables of Abscissas and Weights for Numerical Evaluation of
  Integrals of the Form $\backslash$int{\_}0{\^{}}$\backslash$infty
  e{\^{}}{\{}-x{\}} x{\^{}}n f(x) dx}.
\newblock {\em Mathematical Tables and Other Aids to Computation},
  13(68):285--294, 1959.

\bibitem{Leistedt2012}
B.~Leistedt and J.~D. McEwen.
\newblock {Exact Wavelets on the Ball}.
\newblock {\em IEEE Transactions on Signal Processing}, 60(12):6257--6269,
  2012.

\bibitem{Tang2017a}
Y.-H. Tang.
\newblock {Reference implementation for the algorithms presented in "An
  Atomistic Fingerprint Algorithm for Learning Ab Initio Molecular Force
  Fields"}, 2017.
\newblock DOI: 10.5281/ZENODO.1054550.

\bibitem{Schutt2017}
K.~T. Sch{\"{u}}tt, F.~Arbabzadah, S.~Chmiela, K.~R. M{\"{u}}ller, and
  A.~Tkatchenko.
\newblock {Quantum-chemical insights from deep tensor neural networks}.
\newblock {\em Nature Communications}, 8:13890, 2017.

\bibitem{Chmiela2017}
S.~Chmiela, A.~Tkatchenko, H.~E. Sauceda, I.~Poltavsky, K.~T. Sch{\"{u}}tt, and
  K.-R. M{\"{u}}ller.
\newblock {Machine learning of accurate energy-conserving molecular force
  fields}.
\newblock {\em Science Advances}, 3(5):e1603015, 2017.

\bibitem{Gross2005}
J.~L. Gross and J.~Yellen.
\newblock {\em {Graph theory and its applications}}.
\newblock Chapman {\&} Hall/CRC, 2005.

\bibitem{Liu2003}
M.~Liu, G.~Liu, and K.~Lam.
\newblock {Constructing smoothing functions in smoothed particle hydrodynamics
  with applications}.
\newblock {\em Journal of Computational and Applied Mathematics},
  155(2):263--284, 2003.

\bibitem{L.B.Lucy1977}
{L. B. Lucy}.
\newblock {A numerical approach to the testing of the fission hypothesis}.
\newblock {\em The Astronomical Journal}, 82(12), 1977.

\end{thebibliography}
\bibliographystyle{bibstyle}

\section*{Appendix}

\subsection*{Polynomial Smoothing Functions with Compact Support}

As candidates for the weight of integral and density scaling functions (Section~\ref{sec:radial-weight}), a class of compact polynomials that satisfy the criteria~\cite{Liu2003}:
\begin{enumerate}
\item is compactly supported,
\item is strictly positive within some cutoff distance $r_c$,
\item decreases monotonically,
\item is at least twice continuously differentiable
\end{enumerate}
with minimal number of non-zero terms are:
\begin{align}
W^{a,b}(s) &= \frac{-b\,s^a + a\,s^b}{\sigma},\ a > b > 2, \label{eq:poly-kernel}
\end{align}
where $s = 1 - r / h$ is the normalized complementary coordinate within the span $h$ of the kernel, and
\begin{align}
\sigma = 8 \pi h^3 \left( \frac{a}{b^3+6b^2+11b+6}-\frac{b}{a^3+6a^2+11a+6} \right)
\end{align}
is an optional normalization factor to ensure that the integral of the kernel in a 3D ball of radius $h$ is unity. The parameters $a$ and $b$ are free parameters that can be used to adjust the smoothness and width of the kernel, and can take any real numbers satisfying the condition $a>b>2$. Note that the kernel $W^{4,3}$ is equivalent to the Lucy kernel commonly used in Smoothed Particle Hydrodynamics simulations \cite{L.B.Lucy1977}. The kernel can be evaluated very efficiently using only multiplication and addition when both $a$ and $b$ are integers.

\begin{figure}[htbp!]
	\centering
	\includegraphics[width=0.9\columnwidth]{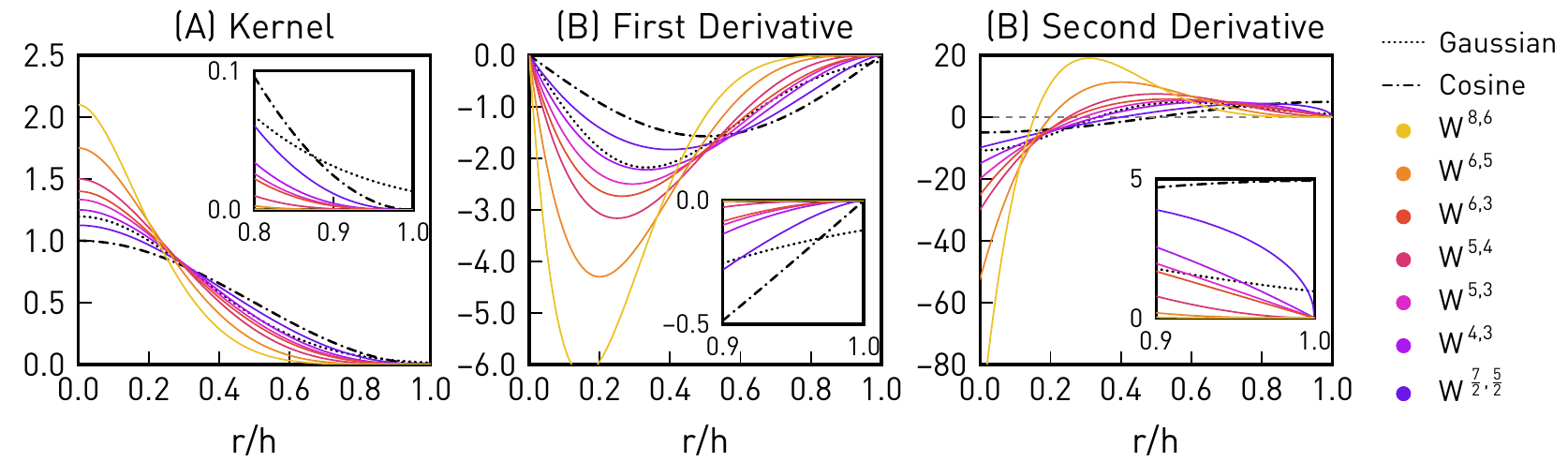}
	\caption{Visualization of the polynomial kernels as given in Eq.~\ref{eq:poly-kernel} with a unit support radius. The kernels are bell-shaped with a derivative of 0 at the origin. Both the first and second derivatives of the kernels transition smoothly to 0 at its support radius. In contrast, the Gaussian kernel and its derivatives does not decay to zero at any finite distance, while the second derivative of the Cosine kernel as mentioned in previous work \cite{Behler2011,Bartok2013a} is not zero at the cutoff distance.\label{fig:kernel}}
\end{figure}

\FloatBarrier

\subsection*{Table of Quadrature Nodes and Weights}

In Table~\ref{table:laguerre-quadrature}, we list the nodes and weights of the Laguerre quadrature rules up to $N_\mathrm{r} = 6$, using notations from Eq.~\ref{eq:laguerre}. In Table~\ref{table:lebedev-quadrature}, we list the nodes and weights of the Lebedev quadrature rules up to $N_\mathrm{r} = 6$, using notations from Eq.~\ref{eq:lebedev}. The Laguerre and Lebedev quadrature nodes can be combined using Eq.~\ref{eq:composite}-\ref{eq:fingerprint-dist-scaled} into composite grids for sampling the atomistic density field.

\begin{table}
\caption{Laguerre quadrature nodes and weights up to 6 points.\label{table:laguerre-quadrature}}
\begin{lstlisting}
================================================================================
                   n = 0      n = 1      n = 2      n = 3      n = 4      n = 5
--------------------------------------------------------------------------------
 Nr = 2    r_n    2.0000     6.0000
           a_n    1.5000     0.5000
--------------------------------------------------------------------------------
 Nr = 3    r_n    1.5174     4.3116     9.1710
           a_n    1.0375     0.9058     0.0568
--------------------------------------------------------------------------------
 Nr = 4    r_n    1.2268     3.4125     6.9027    12.4580
           a_n    0.7255     1.0634     0.2067     0.0044
--------------------------------------------------------------------------------
 Nr = 5    r_n    1.0311     2.8372     5.6203     9.6829    15.8285
           a_n    0.5209     1.0667     0.3835     0.0286     0.0003
--------------------------------------------------------------------------------
 Nr = 6    r_n    0.8899     2.4331     4.7662     8.0483    12.6004    19.2620
           a_n    0.3844     0.9971     0.5361     0.0795     0.0029     0.0000
================================================================================
\end{lstlisting}
\end{table}

\begin{table}
\caption{Lebedev quadrature nodes and weights up to 50 points.\label{table:lebedev-quadrature}}
\begin{lstlisting}
================================================================================
 Na = 6            x_m               y_m               z_m               b_m
 m  = 0,1       +-1.00000           0.00000           0.00000          0.16667
 m  = 2,3         0.00000         +-1.00000           0.00000          0.16667
 m  = 4,5         0.00000           0.00000         +-1.00000          0.16667
--------------------------------------------------------------------------------
 Na = 14           x_m               y_m               z_m               b_m
 m  = 0,1       +-1.00000           0.00000           0.00000          0.06667
 m  = 2,3         0.00000         +-1.00000           0.00000          0.06667
 m  = 4,5         0.00000           0.00000         +-1.00000          0.06667
 m  = 6-13      +-0.57735         +-0.57735         +-0.57735          0.07500
--------------------------------------------------------------------------------
 Na = 26           x_m               y_m               z_m               b_m
 m  = 0,1       +-1.00000           0.00000           0.00000          0.04762
 m  = 2,3         0.00000         +-1.00000           0.00000          0.04762
 m  = 4,5         0.00000           0.00000         +-1.00000          0.04762
 m  = 6-9         0.00000         +-0.70711         +-0.70711          0.03810
 m  = 10-13     +-0.70711           0.00000         +-0.70711          0.03810
 m  = 14-17     +-0.70711         +-0.70711           0.00000          0.03810
 m  = 18-25     +-0.57735         +-0.57735         +-0.57735          0.03214
--------------------------------------------------------------------------------
 Na = 38           x_m               y_m               z_m               b_m
 m  = 0,1       +-1.00000           0.00000           0.00000          0.00952
 m  = 2,3         0.00000         +-1.00000           0.00000          0.00952
 m  = 4,5         0.00000           0.00000         +-1.00000          0.00952
 m  = 6-13      +-0.57735         +-0.57735         +-0.57735          0.03214
 m  = 14-17     +-0.45970         +-0.88807           0.00000          0.02857
 m  = 18-21     +-0.88807         +-0.45970           0.00000          0.02857
 m  = 22-25     +-0.45970           0.00000         +-0.88807          0.02857
 m  = 26-29     +-0.88807           0.00000         +-0.45970          0.02857
 m  = 30-33       0.00000         +-0.45970         +-0.88807          0.02857
 m  = 34-37       0.00000         +-0.88807         +-0.45970          0.02857
--------------------------------------------------------------------------------
 Na = 50           x_m               y_m               z_m               b_m
 m  = 0,1       +-1.00000           0.00000           0.00000          0.01270
 m  = 2,3         0.00000         +-1.00000           0.00000          0.01270
 m  = 4,5         0.00000           0.00000         +-1.00000          0.01270
 m  = 6-9         0.00000         +-0.70711         +-0.70711          0.02257
 m  = 10-13     +-0.70711           0.00000         +-0.70711          0.02257
 m  = 14-17     +-0.70711         +-0.70711           0.00000          0.02257
 m  = 18-25     +-0.57735         +-0.57735         +-0.57735          0.02109
 m  = 26-33     +-0.30151         +-0.30151         +-0.90453          0.02017
 m  = 34-41     +-0.30151         +-0.90453         +-0.30151          0.02017
 m  = 42-49     +-0.90453         +-0.30151         +-0.30151          0.02017
================================================================================
\end{lstlisting}
\end{table}

\end{document}